\documentclass[aps,two column]{revtex4}
\usepackage{color}
\usepackage{psfrag}
\usepackage{dcolumn}
\usepackage{bm}
\usepackage{tikz}
\usepackage[latin1]{inputenc}
\usepackage{graphicx}
\usepackage{amsmath,amssymb}
\usepackage{hyperref}

\begin{document}

\title{Quark stars in massive gravity might be candidates for the mass gap
objects}
\author{J. Sedaghat$^{1}$\footnote{%
J.Sedaghat@shirazu.ac.ir}, B. Eslam Panah$^{2,3,4}$\footnote{%
eslampanah@umz.ac.ir}, R. Moradi$^{5}$\footnote{%
Rahim.Moradi@icranet.org}, S. M. Zebarjad$^{1}$\footnote{%
zebarjad@shirazu.ac.ir}, and G. H. Bordbar$^{1}$ \footnote{%
ghbordbar@shirazu.ac.ir} }
\affiliation{$^{1}$ Department of Physics, Shiraz University, Shiraz 71454, Iran\\
$^{2}$ Department of Theoretical Physics, Faculty of Science, University of
Mazandaran, P. O. Box 47415-416, Babolsar, Iran\\
$^{3}$ ICRANet-Mazandaran, University of Mazandaran, P. O. Box 47415-416,
Babolsar, Iran\\
$^{4}$ ICRANet, Piazza della Repubblica 10, I-65122 Pescara, Italy\\
$^{5}$ Key Laboratory of Particle Astrophysics, Institute of High Energy
Physics, Chinese Academy of Sciences, Beijing 100049, China.}

\begin{abstract}
We have investigated the structural properties of strange quark stars (SQSs)
in a modified theory of gravity known as massive gravity. In order to obtain
the equation of state (EOS) of strange quark matter, we have employed a
modified version of the Nambu-Jona-Lasinio model (MNJL) which includes a
combination of NJL Lagrangian and its Fierz transformation by using
weighting factors ($1-\alpha $) and $\alpha$. Additionally, we have also
calculated dimensionless tidal deformability ($\Lambda$) in massive gravity.
To constrain the allowed values of the parameters appearing in massive
gravity, we have imposed the condition $\Lambda_{1.4 {M}_{\odot
}}\lesssim580 $. Notably, in the MNJL model, the value of $\alpha$ varies
between zero and one. As $\alpha$ increases, the EOS becomes stiffer, and
the value of $\Lambda$ increases accordingy. We have demonstrated that by
softening the EOS with increasing the bag constant, one can obtain objects
in massive gravity that not only satisfy the constraint $\Lambda_{1.4 {M}%
_{\odot }}\lesssim580$, but they also fall within the unknown mass gap
region ($2.5{M}_{\odot}-5{M}_{\odot }$). To establish that the obtained
objects in this region are not black holes, we have calculated Schwarzschild
radius, compactness, and $\Lambda_{{M_{TOV}}}$ in massive gravity.
\end{abstract}

\maketitle

\section{Introduction}

The gravitational wave (GW) events offer new insights into compact stars.
The binary merger GW170817 \cite{Abbott2017} and its electromagnetic
counterpart \cite{Abbott2017b} have led to new constraints on the maximum
mass of neutron stars (NSs). Based on GW170817, the upper bound of $M_{TOV}$
for NSs is predicted as $\sim (2.3-2.4)$$M_{\odot }$ \cite%
{Shibata2019,Rezzolla2018}. However, there are other observations of pulsars
and binary mergers with masses greater than these values such as PSR
J0952-0607 \cite{J0952}, the secondary component of GW190814 \cite%
{Abbott2020,ZMiao2021} and the remnants of GW170817 \cite{Gao2020} and
GW190425 \cite{JSedaghat} which fall within the unknown mass gap region ($2.5%
{M}_{\odot }\lesssim M\lesssim 5{M}_{\odot }$). This region was called a
mass gap because analysis of observations of X-ray binaries revealed a small
number of compact objects in this region \cite{ApJ499,ApJ725,ApJ757}. Recent
observed systems within this interval, show a relative gap instead of an
absolute gap \cite{ApJ941130}. There are arguments that the presence or
absence of mass gap objects can constrain the properties of the
core-collapse supernova engine and the compact objects in the mass gap
region can be created by merging lighter compact objects \cite%
{MNRAS5162022,Drozda2022}. There are two crucial reasons why objects in the
mass gap region are unlikely to be neutron stars. The first reason is that
for non-rotating NSs, the equation of state (EOS) must be very rigid to get
such massive objects (the speed of sound is greater than $0.6$ of the speed
of light) \cite{IBombaci2021,ITews2020}. Such EOSs contrast dimensionless
tidal deformability ($\Lambda $) constraints obtained from GW178017, which
require softer EOSs. The second reason is that for a rotating NS, the
rotation rate should be close to the keplerian limit \cite%
{IBombaci2021,ERMost2020,NBZhang2020,VDexheimer2020}. Now, the question
arises whether these objects are the smallest black holes or other forms of
compact stars, like hybrid stars and self-bounded strange quark stars (SQSs).

Theoretically, strange quark matter (SQM) is formed in two classes of
compact stars. i) Hybrid stars with a quark core and hadronic shells \cite%
{Blaschke2001,Burgio2003,Alford2005,Pal2023,Rather2023,Li2023} and ii) SQSs
composed of pure quark matter from the core up to the upper layers of the
star \cite{Michel1988,Drago2001,Kurkela2010,Wang2019,Deb2021}. Ever since
Trazawa \cite{Terazawa1989}, Witten \cite{Witten1984}, and Bodmer \cite%
{ABodmer} proposed SQM as the ground state of QCD, until now when the GWs
have found so many binaries, the study of SQM in compact stars has always
been of interest. Many studies on hybrid and quark stars have focused on
recent discoveries of GWs. Refs. \cite{Daniel2021,AngLi2021,Ferreira2021}
study hybrid stars with constraints obtained from GW190814, GW170817, and
PSR J0030+0451 \cite{Miller2019,Riley2019}. In Ref. \cite{ShuHua2020}, the
properties of SQM in the bag model and non-newton gravity are investigated
by the constraints obtained from GW170817 and PSR J0740+6620. In Ref. \cite%
{ALi2021}, the EOS of quark matter is constrained by using the bayesian
statistical model and the mass and radius measurements of PSR J0030+0451
\cite{TERiley2019}. Refs. \cite{IBombaci2021,ZMiao2021,Roupas2021}
investigate whether the secondary component of GW190814 could be a SQS. In
Refs. \cite{Lopes2021,CZhang2021} the structural features of the SQSs have
also been investigated using the data obtained from GWs.

GR is a successful theory of gravity. However, at large scales, it cannot
explain why our universe is undergoing an accelerated cosmic expansion. This
is one of the reasons why GR needs to be modified. Among various modified
theories of gravity, massive gravity can describe the late-time acceleration
without considering dark energy \cite{MassI,MassII,MassIII,MassIV,MassV}.
This theory of gravity modifies gravitational effects by weakening them at a
large scale compared to GR, which allows the universe to accelerate.
However, its predictions at small scales seem to be the same as those of GR.
Massive gravity will result in the graviton having a mass of $m_{g}$, which
in the absence of the mass of the graviton (i.e., $m_{g}\rightarrow 0$),
this theory reduces to GR. In Ref. \cite{Zhang2019}, it has been indicated
that the mass of a graviton is very small in the usual weak gravity
environments but becomes much larger in the strong gravity regimes, such as
black holes and other compact objects. In addition, recent observations by
the advanced LIGO/Virgo collaborations have put a tight bound on the
graviton mass \cite{LIGOI,LIGOII}. Also, other theoretical and empirical
limits on the graviton's mass exist \cite%
{massgraviton1,massgraviton2,massgraviton3,massgraviton4}. In the
astrophysics context and by considering massive gravity, the properties of
compact objects such as relativistic stars \cite{Kareeso2018,Yamazaki2019},
neutron stars \cite{Katsuragawa2016,Behzad2017}, and white dwarfs \cite%
{Behzad2019}, have been studied. In 2011, de Rham, Gabadadze and Tolley
(dRGT) introduced a ghost-free theory of massive gravity \cite%
{deRham,Hinterbichler2012} which is known dRGT massive gravity. This theory
uses a reference metric to construct massive terms (see Refs. \cite%
{deRham,Hinterbichler2012,Hassan2012}, for more details). These massive
terms are inserted in the action to provide massive gravitons. Then, Vegh
extended the dRGT massive gravity theory using holographic principles and a
singular reference metric in 2013 \cite{Vegh}, which is known as dRGT-like
massive gravity. Recently, this extended theory of dRGT massive gravity
attracted many people in various viewpoints of physics.

In this paper, we have used a modified version of the Nambu-Jona-Lasinio\
(MNJL) model to obtain the structural properties of SQS in massive gravity.
In GR, the soft EOSs give small masses for quark stars, and the stiff EOSs
do not satisfy the $\Lambda $ constraint. We show that in massive gravity,
it is possible to have quark stars that not only have masses higher than
those in GR, but also satisfy the GW observational constraints well. We have
used $\Lambda $ constraint ($\Lambda _{1.4{M}_{\odot }}\lesssim580$) \cite%
{Abbott2018} to restrict the masses of SQSs in massive gravity.

\section{Modified NJL Model}

\label{mnjl} We have used a modified self-consistent version of the $(2+1)-$%
flavor NJL model in mean field approximation \cite{ChengMingLi2020} to
derive the EOS. The modified Lagrangian is a combination of the original NJL
Lagrangian and its Fierz transformation \cite{Hatsuda1985} by using
weighting factors $(1-\alpha )$ and $\alpha $ as follows \cite%
{ChengMingLi2020}
\begin{align}
\mathcal{L}& =(1-\alpha )\mathcal{L}_{NJL}+\alpha \mathcal{L}_{F}  \notag \\
& =\bar{\psi}(i{\partial\!\!\!\big /}-m)\psi +(1-\alpha )G\sum_{i=0}^{8}\left[ (\bar{%
\psi}\lambda _{i}\psi )^{2}+(\bar{\psi}i\gamma ^{5}\lambda ^{i}\psi )^{2}%
\right]  \notag \\
& -\frac{\alpha G}{2}\left[ \left( \bar{\psi}\gamma ^{\mu }\lambda
_{0}^{C}\psi \right) ^{2}-\left( \bar{\psi}\gamma ^{\mu }\gamma ^{5}\lambda
_{0}^{C}\psi \right) ^{2}\right]  \notag \\
& -K\,\left( \det [\bar{\psi}(1+\gamma ^{5})\psi ]+\det [\bar{\psi}(1-\gamma
^{5})\psi ]\right) ,  \label{lagrangiantotal}
\end{align}%
where $\mathcal{L}_{NJL}$ is the original Lagrangian in flavor space without
Fierz transformation and $\mathcal{L}_{F}$ is the Fierz transformation of $%
\mathcal{L}_{NJL}$ in color space \cite{ChengMingLi2020}. $\psi $ denotes a
quark field with three flavor (u, d, s) and three color degrees of freedom. $%
m$ is the corresponding mass matrix in flavor space. In addition, $\gamma
^{\mu }$s ($\mu =0\longrightarrow 3$) and $\lambda ^{i}$s ($%
i=1\longrightarrow 8$) are Dirac and Gell-Mann matrices, respectively. $%
\gamma ^{5}=i\gamma ^{0}\gamma ^{1}\gamma ^{2}\gamma ^{3}$ and $\lambda ^{0}=%
\sqrt{\dfrac{2}{3}I}$, where $I$ is the identity matrix. Similar to $\lambda
^{0}$, $\lambda _{0}^{C}$ is equal to $\sqrt{\dfrac{2}{3}I}$ but in color
space \cite{ChengMingLi2020}. Also, $G$ and $K$ are the four-fermion and
six-fermion interaction coupling constants, respectively. The parameter $%
\alpha $ varies from zero to one. By setting $\alpha =0$, the Lagrangian \ref%
{lagrangiantotal} converts to original NJL Lagrangian and for $\alpha =1$,
it converts to Fierz transformation of $\mathcal{L}_{NJL}$. In Ref. \cite%
{ChengMingLi2020}, the authors have used the Lagrangian (\ref%
{lagrangiantotal}) and have shown that this Lagrangian gives the EOSs with $%
M_{TOV}\simeq 2{M}_{\odot }$ in GR which satisfy structural constraints
(mass, radius and $\Lambda $) obtained by GWs \cite{ChengMingLi2020}. They
have used the first analysis of $\Lambda $ with the upper bound of $\Lambda
_{1.4{M}_{\odot }}\leq 800$ \cite{Abbott2017} and have obtained a SQS with $%
M_{TOV}\simeq 2.057{M}_{\odot }$ and $\Lambda _{1.4{M}_{\odot }}=634$. In
this paper, we show that by using the Lagrangian (\ref{lagrangiantotal}) and
massive gravity, SQSs can be obtained with masses in gap region ($2.5{M}%
_{\odot }\lesssim M\lesssim 5{M}_{\odot }$) and simultaneously, respect to
the improved analysis of $\Lambda $ with the upper bound of $\Lambda _{1.4{M}%
_{\odot }}\lesssim 580$ \cite{Abbott2018}.

\begin{figure}[h]
\center{\includegraphics[width=7.5cm] {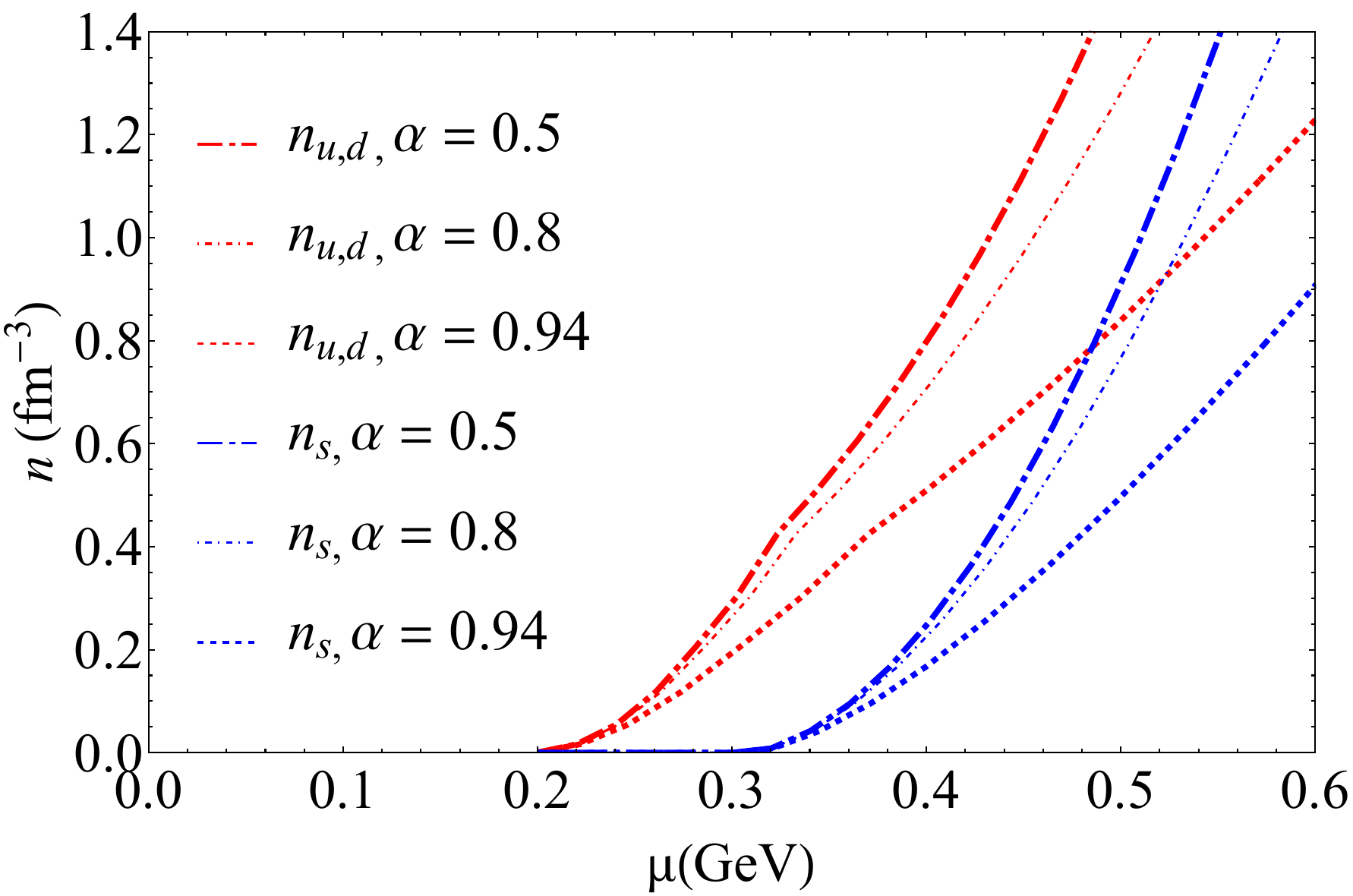}}
\caption{Quark number density ($n$) versus chemical potential ($\protect\mu $%
). Notably, $n_{u,d}$ devotes to up and down quark number densities, and $%
n_{s}$ is related to the strange quark number density.}
\label{density of quarks}
\end{figure}

In the following, we use the Lagrangian (\ref{lagrangiantotal}) and obtain
the EOS of SQM in mean field approximation. As the first step, we calculate
the quark number densities for different choices of $\alpha $ (to know the
details of calculations, see the Ref. \cite{ChengMingLi2020}). Fig. \ref%
{density of quarks} shows quark number densities of up, down, and strange
flavors as a function of chemical potential for different values of $\alpha $%
. As one can see from the figure \ref{density of quarks}, the quark number
densities of up and down quarks are zero for $\mu \lesssim 200MeV$, and the
strange quark number density is zero for $\mu \lesssim 320MeV$. This figure
shows that the quark number densities decrease with increasing $\alpha $.
Consequently, such behavior also occurs for baryon number density (for more
discussions, see Ref. \cite{ChengMingLi2020}).

The following conditions should be imposed to obtain the EOS of SQM, which
are:
\begin{align}
& \mu _{s}=\mu _{d}\equiv \mu ,~~~\&~~~\mu _{u}=\mu -\mu _{e},  \notag \\
&  \notag \\
& \frac{2}{3}n_{u}-\frac{1}{3}n_{d}-\frac{1}{3}n_{s}-n_{e}=0,  \notag \\
&  \notag \\
& n_{B}=\dfrac{n_{u}+n_{d}+n_{s}}{3}.  \label{SQM conditions}
\end{align}%
The first two lines of Eq. (\ref{SQM conditions}) come from beta equilibrium
and charge neutrality conditions, respectively. In the above conditions, $%
\mu _{i}$s $(i=u,d,s)$ are quark chemical potentials with flavor $i$, and $%
\mu _{e}$ is the electron chemical potential. $n_{i}$s $(i=u,d,s)$ are quark
number densities with flavor $i$. In addition, $n_{B}$ is the baryon number
density. Furthermore, $n_{e}=\mu _{e}^{3}/(3\pi ^{2})$ is the electron
number density. Then the pressure and the energy density can be obtained by
the following relations
\begin{align}
& P=P_{0}+\int_{0}^{\mu _{j}}d\mu _{j}^{^{\prime }}\ n_{j}(\mu
_{j}^{^{\prime }}),\ \ \ \ \ \ \ (j=u,d,s,e),  \notag \\
&  \notag \\
& \epsilon =-P+\mu _{j}\ n_{j}(\mu _{j}),
\end{align}%
where $P_{0}$ is the vacuum pressure, which is taken as the vacuum bag
constant ($-B$). We consider two values for $B$ including $B^{1/4}=117MeV$
and $B^{1/4}=130MeV$. For these values of $B$, the energy per baryon of
three-flavor quark matter is lower than that of two-flavor quark matter for $%
\alpha \leq 0.94$ \cite{ChengMingLi2020}. This condition is necessary for
SQM to be the true ground state of the strong interaction \cite{weber 2005}.
In the following, we present our results for the structural properties of
SQS in massive gravity for different values of $B$ and $\alpha $. To this
end, we first explain the modified TOV equation in massive gravity and then
derive the differential equations to calculate the $\Lambda $ in this theory
of gravity.

\section{Modified TOV Equation in Massive Gravity}

Here, we consider dRGT-like massive gravity, which is introduced in Ref.
\cite{Vegh}. Notably, the dRGT-like massive gravity is similar to the dRGT
massive theory of gravity, which is proposed by de Rham-Gabadadze-Tolley
(dRGT) in 2011 \cite{deRham}. The dRGT-like massive gravity's action is
given by \cite{Vegh}
\begin{equation}
\mathcal{I}=\frac{1}{2\kappa^2 }\int d^{4}x\sqrt{-g}\left[ \mathcal{R}%
+m_{g}^{2}\sum_{i}^{4}c_{i}\mathcal{U}_{i}(g,f)\right] +\mathcal{I}_{matter},
\label{Action}
\end{equation}%
where $\mathcal{R}$ and $m_{g}$ are the Ricci scalar and the graviton mass,
respectively. It is notable that $\kappa^2 =$\ $8\pi $ because we consider $%
G=c=1$. Also, $f$ and $g$ are fixed symmetric and metric tensors,
respectively. In the above action, $\mathcal{I}_{matter}$ is related to the
action of matter. Moreover, $c_{i}$'s are free parameters of the massive
theory, which are arbitrary constants. Notabley, their values can be
determined according to theoretical or observational considerations \cite%
{Amico,Akrami,Berezhiani}. $\mathcal{U}_{i}$'s are symmetric polynomials of
the eigenvalues of $4\times 4$ matrix $K_{\nu}^{\mu}=\sqrt{%
g^{\mu\alpha}f_{\alpha \nu }}$ (where $g_{\mu \nu }$ is the dynamical
(physical) metric, and $f_{\mu \nu }$ is the auxiliary reference metric,
needed to define the mass term for gravitons) where they can be written in
the following form
\begin{equation}
\mathcal{U}_{i}=\sum_{y=1}^{i}\left( -1\right) ^{y+1}\frac{\left( i-1\right)
!}{\left( i-y\right) !}\mathcal{U}_{i-y}\left[ \mathcal{K}^{y}\right],
\end{equation}%
where $\mathcal{U}_{i-y}=1$, when $i=y$. Also, $\mathcal{K}$\ stands for
matrix square root, i.e., $\mathcal{K}_{~~~\nu }^{\mu }=\left( \sqrt{%
\mathcal{K}}\right) _{~~~\lambda }^{\mu }\left( \sqrt{\mathcal{K}}\right)
_{~~~\upsilon }^{\lambda }$, and the rectangular bracket denotes the trace $%
\left[ \mathcal{K}\right] =\mathcal{K}_{~~~\mu }^{\mu }$.

We consider the four-dimensional static spherical symmetric spacetime
\begin{equation}
ds^{2}=e^{2\Phi(r)}dt^{2}-\frac{dr^{2}}{e^{-2\lambda(r)}}%
-r^{2}h_{ij}dx_{i}dx_{j},~\text{where}~i,j=1,2.  \label{Metric}
\end{equation}%
where $e^{2\Phi(r)}$ and $e^{-2\lambda(r)}$ are unknown metric functions,
and $h_{ij}dx_{i}dx_{j}=\left( d\theta ^{2}+\sin ^{2}\theta d\varphi
^{2}\right) $. The equation of motion of dRGT-like massive gravity is given
in the following form
\begin{equation}
G_{\mu }^{\nu }+m_{g}^{2}\chi _{\mu }^{\upsilon }=8\pi T_{\mu }^{\nu },
\label{field1}
\end{equation}%
where $G_{\mu }^{\nu }$ is the Einstein tensor, and $T_{\mu }^{\nu }$
denotes the energy-momentum tensor which comes from the variation of $%
\mathcal{I}_{matter}$ and $\chi _{\mu \nu }$ is the massive term with the
following explicit form
\begin{equation}
\chi _{\mu \nu }=-\sum_{i=1}^{d-2}\frac{c_{i}}{2}\left[ \mathcal{U}%
_{i}g_{\mu \nu }+\sum_{y=1}^{i}\left( -1\right) ^{y}\frac{\left( i!\right)
\mathcal{U}_{i-y}\left[ \mathcal{K}_{\mu \nu }^{y}\right] }{\left(
i-y\right) !}\right] .  \notag
\end{equation}

Considering the quark stars as a perfect fluid, the nonzero components of
the energy-momentum tensor are given by \cite{Behzad2017}
\begin{equation}
T_{0}^{0}=\epsilon ,~~~T_{1}^{1}=T_{2}^{2}=T_{3}^{3}=-P.
\end{equation}

We follow the reference metric $f_{\mu \nu }$\ for four-dimensional static
spherical symmetric spacetime as it is considered for studying neutron stars
and white dwarfs in the following form \cite{Behzad2017,Behzad2019}
\begin{equation}
f_{\mu \nu }=diag(0,0,-J^{2},-J^{2}\sin ^{2}\theta ),  \label{referenceM}
\end{equation}%
in which $J$\ is known as a parameter of the reference metric, which is a
positive constant. The explicit functional forms of $U_{i}$'s are given by
\cite{Behzad2017}
\begin{eqnarray}
\mathcal{U}_{1} &=&\frac{2J}{r},  \notag \\
&&  \notag \\
\mathcal{U}_{2} &=&\frac{2J^{2}}{r^{2}},  \notag \\
&&  \notag \\
\mathcal{U}_{i} &=&0,~~~~~i>2.
\end{eqnarray}

Using the field equation and reference and dynamical metrics, the metric
function $e^{-2\lambda (r)}$, is obtained in the following form \cite%
{Behzad2017}
\begin{equation}
e^{-2\lambda (r)}=1-m_{g}^{2}J\left( \frac{c_{1}r}{2}+c_{2}J\right) -\frac{%
2M(r)}{r},  \label{4g(r)}
\end{equation}%
in which $M(r)=\int_{0}^{R}4\pi r^{2}\epsilon (r)dr$ $\left( \text{or }\frac{%
dM(r)}{dr}=4\pi r^{2}\epsilon (r)\,\right) $. By definition $C_{1}\equiv
m_{g}^{2}c_{1}J$ and $C_{2}\equiv m_{g}^{2}c_{2}J^{2}$ in Eq. (\ref{4g(r)}),
the modified TOV equation in massive gravity is given by \cite{Behzad2017}
\begin{equation}
\frac{dP(r)}{dr}=\frac{\left( M(r)+4\pi r^{3}P-\frac{r^{2}C_{1}}{4}\right)
\left( \epsilon +P\right) }{r\left( \frac{C_{1}r^{2}}{2}+2M(r)+r\left(
C_{2}-1\right) \right) }.\,\,  \label{TOVm2}
\end{equation}

Here, we obtain $\Lambda $ in massive gravity. To do this, we need to get a
differential equation of metric function ($H(r)$) in massive gravity. In
Appendix A, we have shown how we obtained this equation in massive gravity.
Also, to check the correctness of the calculations, we have investigated the
limit $C_{1}=C_{2}=0$ and shown that in this limit, the differential
equation obtained in massive gravity is converted to the differential
equation in GR. The differential equation to get $H(r)$ in massive gravity
is
\begin{equation}
\left( A_{1}+A_{2}+A_{3}\right) H(r)+A_{4}\beta +\frac{d\beta }{dr}=0,
\label{HbetaEq}
\end{equation}%
where $\beta =dH(r)/dr$, and $A_{i}$s are as follows
\begin{eqnarray}
A_{1} &=&\frac{6}{r^{2}B_{1}},  \notag \\
&&  \notag \\
A_{2} &=&\frac{B_{1}}{2r^{2}B_{1}^{2}}\left\{ \frac{B_{2}\left(
B_{2}f+B_{3}\right) }{B_{1}}-7B_{2}+3B_{3}\right.  \notag \\
&&+\frac{1}{B_{4}}\left[ \frac{(f+1)B_{2}}{\left( B_{2}+\frac{C_{1}r}{2}%
\right) ^{-1}}+\frac{2B_{1}\left( B_{2}+\frac{C_{1}r}{2}\right) }{%
B_{4}\left( \frac{2M\left( r\right) }{r}+8\pi r^{2}\epsilon -2\right) ^{-1}}%
\right.  \notag \\
&&+16\pi r^{2}\left[ \epsilon \left( \frac{C_{1}r}{4}+C_{2}+4\pi
pr^{2}-1\right) \right.  \notag \\
&&\left. \left. \left. +\frac{3p}{\left( \frac{5C_{1}r}{12}+C_{2}+\frac{4\pi
pr^{2}}{3}-1\right) ^{-1}}+\frac{M\left( r\right) (7p+3\epsilon )}{r}\right] %
\right] \right\} ,  \notag \\
&&  \notag \\
A_{3} &=&\frac{4\pi (p+\epsilon )f}{B_{1}},  \notag \\
&&  \notag \\
A_{4} &=&\frac{2\left( \frac{3C_{1}r}{4}+2\pi r^{2}(\epsilon
-p)+C_{2}-1\right) +\frac{2M\left( r\right) }{r}}{rB_{1}},
\end{eqnarray}%
where $B_{1}$, $B_{2}$, $B_{3}$, and $B_{4}$ are
\begin{eqnarray}
B_{1} &=&\frac{C_{1}r}{2}+C_{2}-1+\frac{2M\left( r\right) }{r},  \notag \\
&&  \notag \\
B_{2} &=&8\pi pr^{2}-\frac{C_{1}r}{2}+\frac{2M\left( r\right) }{r},  \notag
\\
&&  \notag \\
B_{3} &=&\frac{2M\left( r\right) }{r}-\frac{C_{1}r}{2}-8\pi \epsilon r^{2},
\notag \\
&&  \notag \\
B_{4} &=&1-\frac{2M\left( r\right) }{r},
\end{eqnarray}%
also, $f$ is defined as $d\epsilon /dP$. In addition, $\Lambda $ is then
obtained by
\begin{equation}
\Lambda =\frac{2}{3}k_{2}R^{5},
\end{equation}%
where $k_{2}$ is dimensionless tidal Love number for $l=2$, which is given
as
\begin{align}
k_{2}& =\frac{16\sigma ^{5}}{5}(1-2\sigma )^{2}\left[ 1+\sigma (y-1)-\frac{y%
}{2}\right]  \notag \\
& \times \left\{ 12\sigma \left[ 1-\frac{y}{2}+\frac{\sigma (5y-8)}{2}\right]
\right.  \notag \\
& +4\sigma ^{3}\left[ 13-11y+\sigma \left( 3y-2\right) +2\sigma ^{2}\left(
1+y\right) \right]  \notag \\
& +\left. 6(1-2\sigma )^{2}\left[ 1+\sigma (y-1)-\frac{y}{2}\right] \ln
\left( 1-2\sigma \right) \right\} ^{-1/2},  \label{tln}
\end{align}%
where $\sigma =M/R$ and $y$ is defined as $y=R\beta (R)/H(R)-4\pi
R^{3}\epsilon _{0}/M$ in which $\epsilon _{0}$ is the energy density at the
surface of the quark star \cite{ChengMingLi2020}. By simultaneously solving
equations (\ref{TOVm2}) and (\ref{HbetaEq}) along with $dM/dr=4\pi
r^{2}\epsilon $, we obtain the curves of star mass in terms of radius and $%
\Lambda $ in terms of mass. The results are presented in the next section.

\subsection{Modified Schwarzschild Radius}

It is clear that by considering dRGT-like massive gravity, the Schwarzschild
radius is modified. Using Eq. (\ref{4g(r)}), we can get the Schwarzschild
radius ($R_{Sch}$) for this theory of gravity. To find the modified
Schwarzschild radius, we solve $S(r)=0$. So, the modified Schwarzschild
radius for dRGT-like massive gravity is given \cite{Behzad2017}
\begin{equation}
R_{Sch}=\frac{\left( 1-C_{2}\right) }{C_{1}}-\frac{\sqrt{\left(
C_{2}-1\right) ^{2}-4C_{1}M}}{C_{1}},  \label{Sch}
\end{equation}%
where $M=M\left(r=R\right)$ is the star's total mass (or the mass of the star on its surface).

\subsection{Compactness}

The compactness of a spherical object may be defined by the ratio of mass to
radius of that object
\begin{equation}
\sigma =\frac{M}{R},
\end{equation}%
which may be indicated as the strength of gravity. In the following, we
calculate the thermodynamic properties of SQM (EOS, speed of sound, and
adiabatic index) for different values of $\alpha $ and $B$. We investigate
the structural properties of SQS (mass, radius, dimensionless tidal
deformability, compactness, and Schwarzschild radius) for a stiff EOS ($%
B^{1/4}=117MeV$), and a soft EOS ($B^{1/4}=130MeV$) in massive gravity. The
value of $\Lambda $ depends not only on the EOS but also on the gravity
used. In GR, the stiff EOSs cause the $\Lambda $ constraint to be lost and
the soft EOSs can not cover the objects fall within in the mass gap region.
We demonstrate that in massive gravity, the soft EOSs can not only lead to
masses in the mass gap region but also satisfy the $\Lambda $ constraint
well.

\section{Equation of States of SQM for $B^{1/4}=117MeV$}

\label{eos-1}

\begin{figure}[h]
\center{\includegraphics[width=7.2cm] {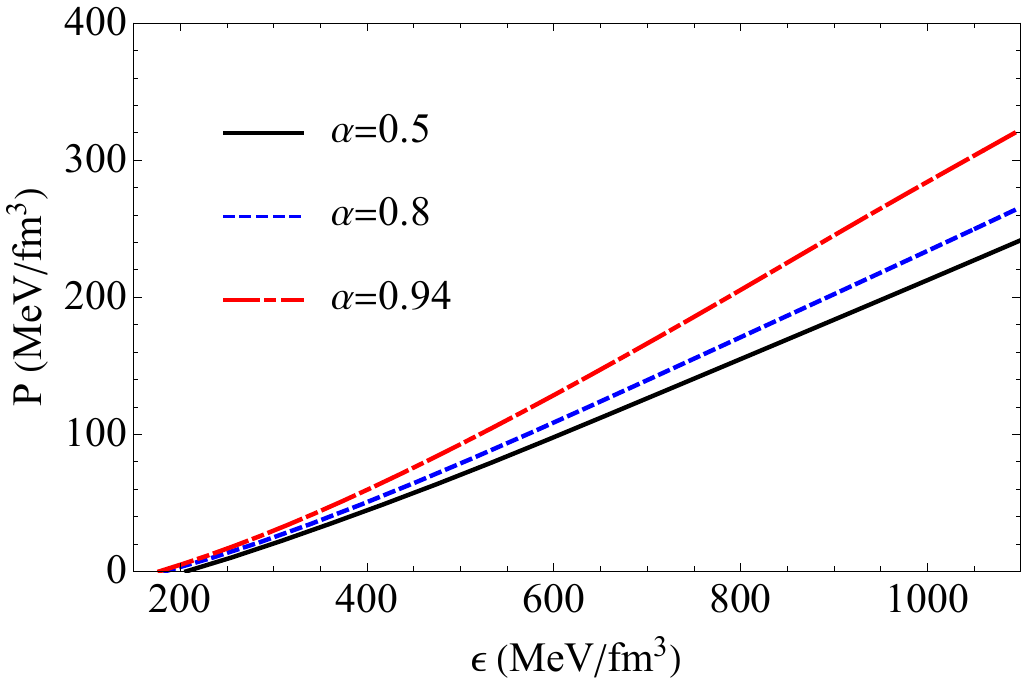}}
\caption{EOSs of SQM for $B^{1/4}=117MeV$ and different values of $\protect%
\alpha $.}
\label{eos-njl}
\end{figure}

\begin{figure}[h]
\center{\includegraphics[width=7.5cm] {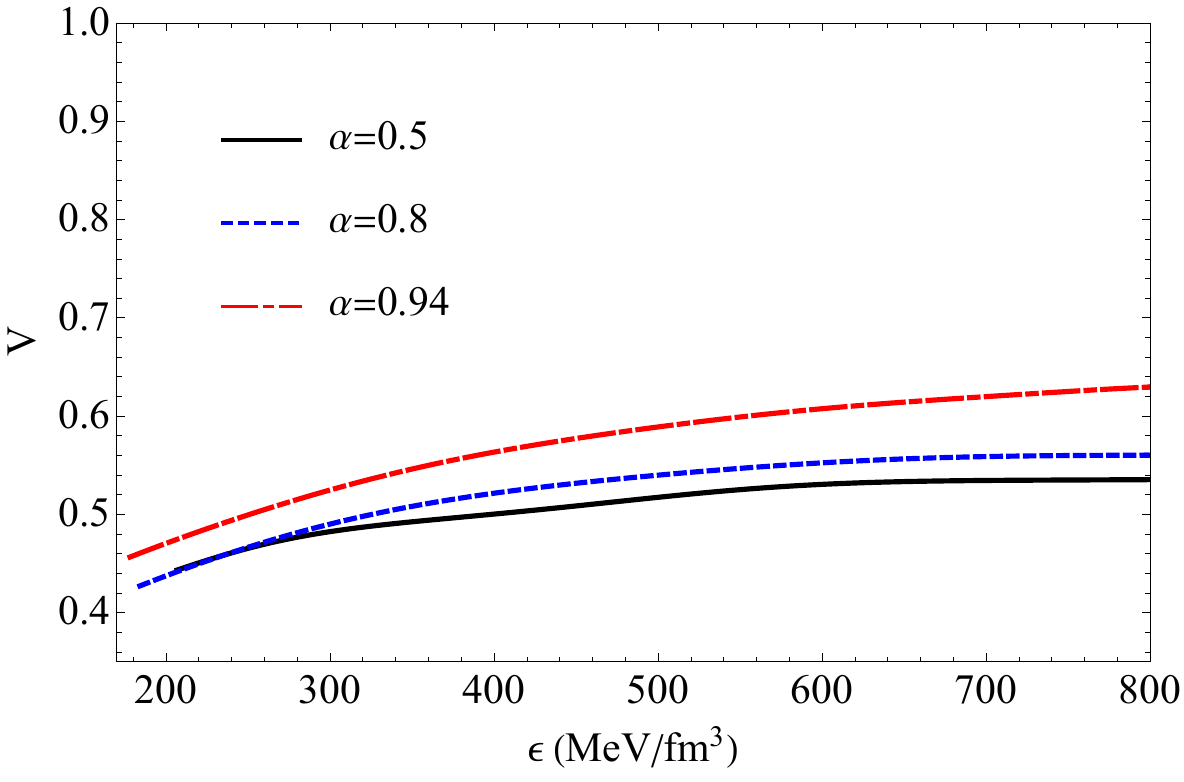}}
\caption{The speed of sound ($V$) energy density ($\protect\epsilon $) for
different values of $\protect\alpha $.}
\label{sound-speednjl}
\end{figure}

In this section, we present our results for the thermodynamic properties of
SQM in a modified NJL model for $B^{1/4}=117MeV$ and different values of $%
\alpha $. As it was mentioned in section \ref{mnjl}, by choosing this value
of bag constant, the parameter $\alpha $ should be equal or less than $0.94$
to establish the stability condition of SQM. We derive the EOSs of SQM for
different values of $\alpha $ including $\alpha =0.5$, $\alpha =0.8$, and $%
\alpha =0.94$. The results are presented in Fig. \ref{eos-njl}. This figure
shows that the EOSs become stiffer by increasing the $\alpha $ parameter. We
will show in section \ref{structural properties} that this behavior causes
the mass of the star to increase. But it is possible that the structural
features of these stars do not satisfy the constraints obtained from GW
observations. Therefore, at the same time as the mass of the star increases,
such constraints should also be checked. To check the causality, the speed
of sound in terms of energy density is plotted in Fig. \ref{sound-speednjl}.
The plots indicate that causality is established well for all $\alpha $
values. Also, with an increase in the value of $\alpha $, the speed of sound
increases, which means that the EOS becomes stiffer and subsequently, the
mass of the star becomes greater.

\begin{figure}[h]
\center{\includegraphics[width=7.5cm] {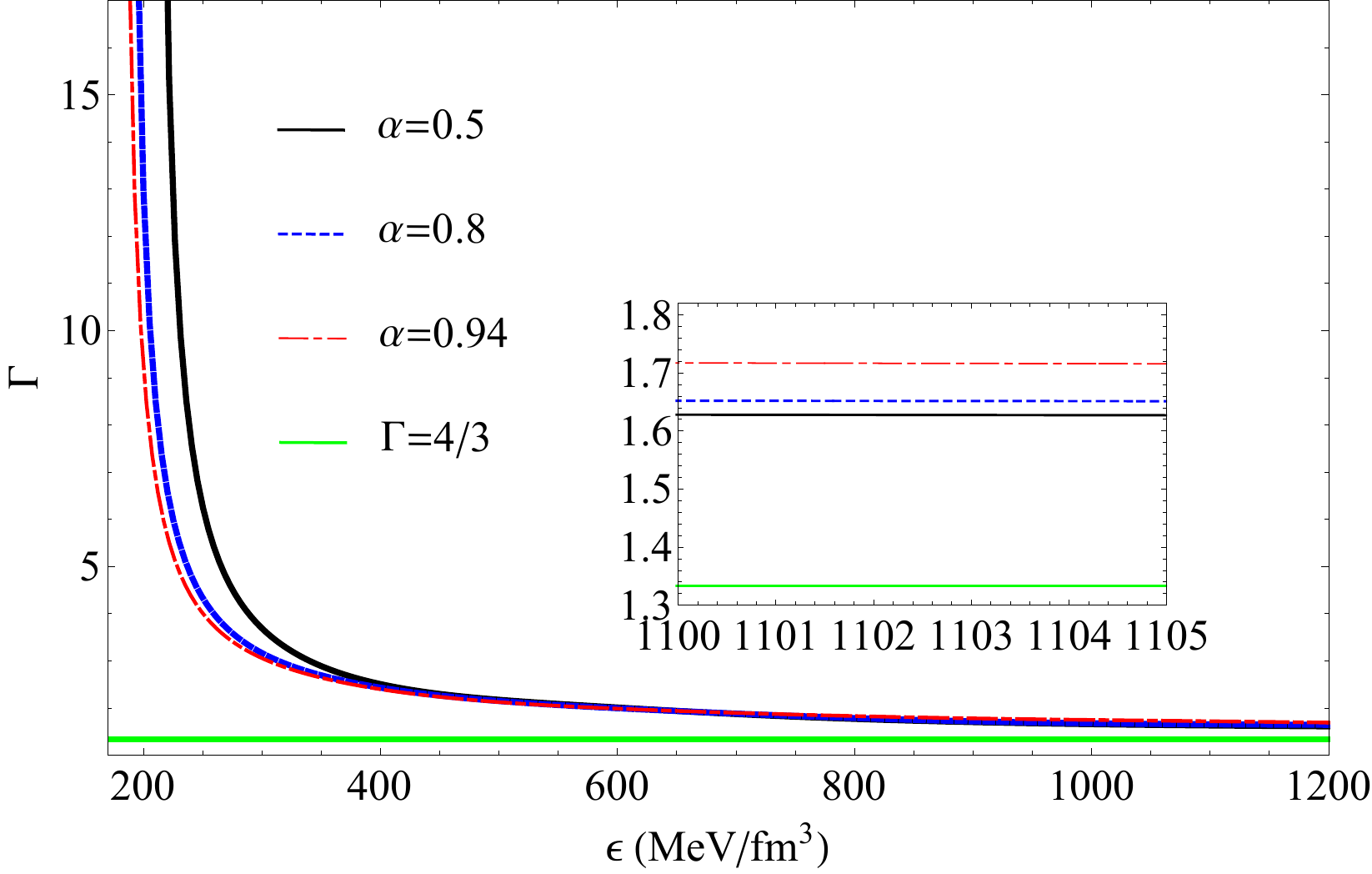}}
\caption{Adiabatic index ($\Gamma $) versus energy density ($\protect%
\epsilon $) for $B^{1/4}=117MeV$ and different values of $\protect\alpha $.}
\label{adianjl}
\end{figure}

As a measure of dynamical stability, the adiabatic index $\left( \Gamma =%
\frac{dP}{d\epsilon }\frac{P+\epsilon }{P}\right)$ has been presented as a
function of energy density in Fig. \ref{adianjl}. It is known that $\Gamma $
should be higher than $4/3$ for dynamical stability \cite%
{Behzad2017,Knutsen1988,Mak2013,Sedaghat20221,Hendi2016}. As one can see
from this figure, this condition is satisfied for all values of $\alpha $.

\subsection{Structural Properties of SQS in Massive Gravity for $%
B^{1/4}=117MeV$}

\label{structural properties} In this section, we obtain the structural
properties of SQS, including the $M-R$ and $\Lambda -M$ relations in
dRGT-like massive gravity. We then compare the results with the constraints
derived from astronomical observations. In addition, we obtain modified
Schwarzschild radius and compactness of quark stars in this theory of
gravity.

Practically, the modified TOV equation (\ref{TOVm2}) is solved simultaneously with the equation, $M(r)=\int_{0}^{R}4\pi r^{2}\epsilon(r)dr $. To solve these equations, we have to consider the modified NJL EoS with the obtained pressure at the center of the star (i.e. $P(r=0)=P_{c}$, which $P_{c}$ is the central pressure of the quark star and given by the modified NJL EoS). Notably, the mass of the star at the center of the star is zero, i.e. $M(r=0)=0$. Then, we continue to solve the equations up to the star's surface, where the pressure is zero ($P(r=R)=0$). The result is the mass and corresponding radius of the star (i.e. $M=M(R)$). We do this for different values of central pressures allowed by the mentioned EOS.

\subsubsection{$\protect\alpha =0.5$}

\begin{figure}[h]
\center{\includegraphics[width=8.5cm] {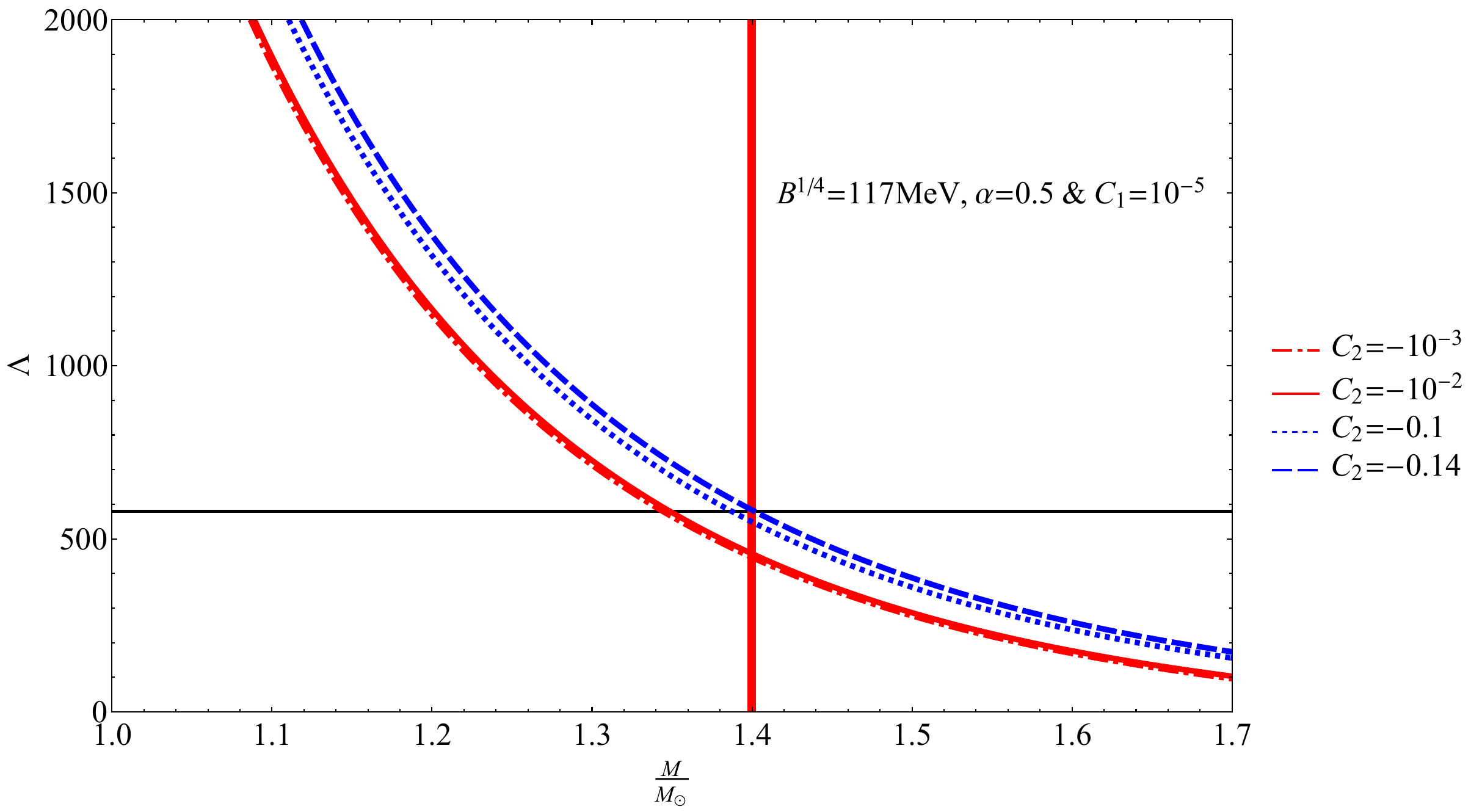}}
\caption{Dimensionless tidal deformability ($\Lambda $) versus mass of the
star for $B^{1/4}=117MeV,\protect\alpha =0.5,C_{1}=10^{-5}$ and different
values of $C_{2}$. The horizontal black line and the vertical red line
correspond to $\Lambda =580$ and $\frac{M}{{M}_{\odot }}=1.4$, respectively.}
\label{tidal-alph=0.5-data1}
\end{figure}

Now, we investigate the structural properties of SQS in massive gravity. We
first use the obtained EOSs in section \ref{eos-1} and the modified TOV
equation (\ref{TOVm2}) to derive the mass-radius ($M-R$) relation and $%
\Lambda -M$ relation in massive gravity. As we saw in section \ref{eos-1},
we used three different values of $\alpha $ including $0.5$, $0.8$, and $%
0.94 $ for the EOSs in the MNJL model. Now, we first examine the case $%
\alpha =0.5 $. The equation (\ref{TOVm2}) shows that there are two
parameters, including $C_{1}$ and $C_{2}$, in massive gravity (these
parameters are related to the dRGT-like massive gravity's parameters). In
the beginning, we set a fixed value for $C_{1}$ and investigate the
structural features of SQS for different values of $C_{2}$. Figures \ref%
{massnjl1} and \ref{tidal-alph=0.5-data1} show the $M-R$ and $\Lambda -R$
diagrams, respectively, for $\alpha =0.5$, $C_{1}=10^{-5}$, and different
values of $C_{2}$.
\begin{figure}[h]
\center{\includegraphics[width=8.5cm] {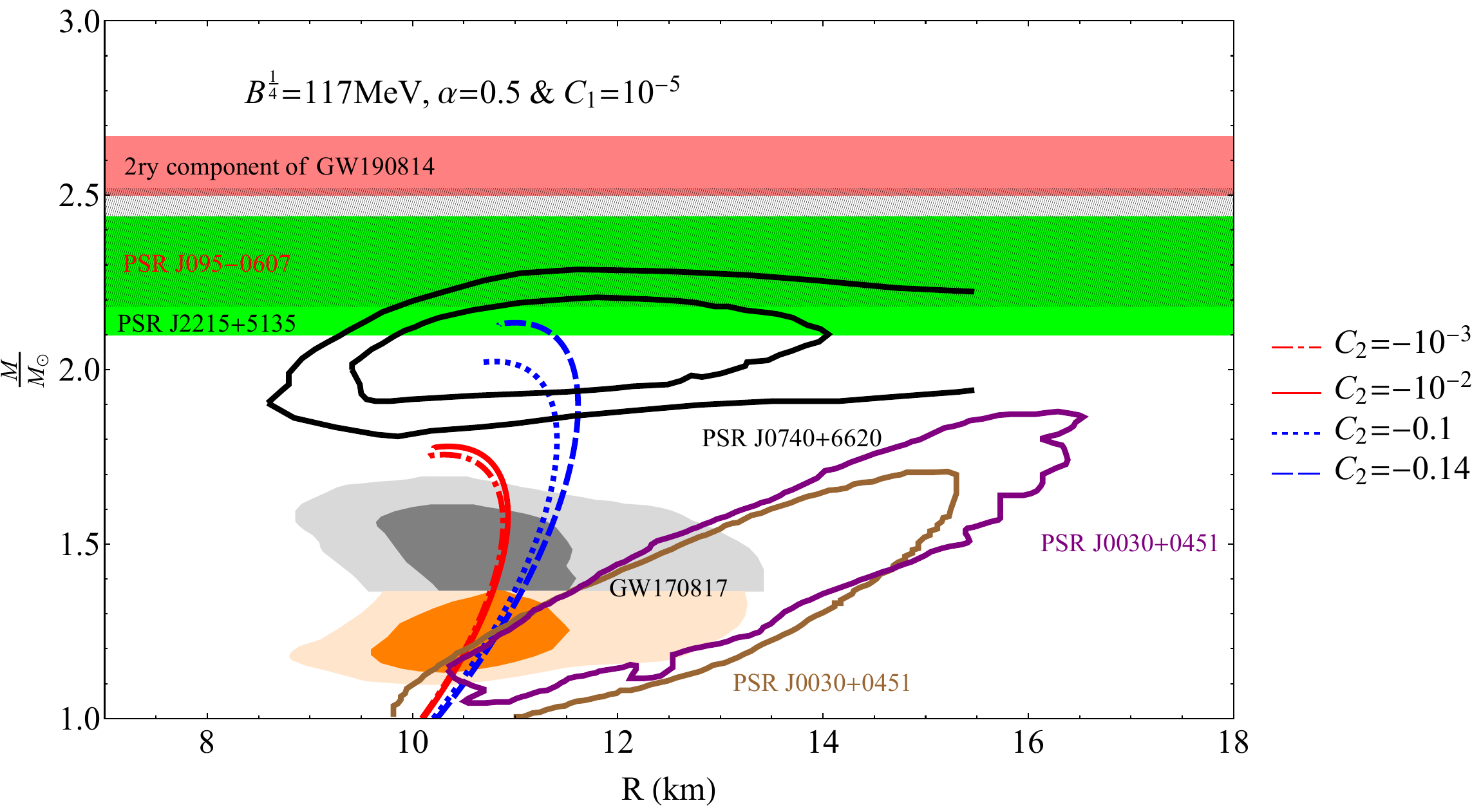}}
\caption{Mass-Radius diagram for $B^{1/4}=117MeV$, $\protect\alpha =0.5,%
\hspace{1mm}C_{1}=10^{-5}$ and different values of $C_{2}$. The gray and
orange regions are the mass - radius constraints from the GW170817 event.
The black region shows pulsar J0740+6620, The green and black hatched
regions represent pulsars J2215+5135 \protect\cite{Linares2018} and PSR
J095-0607, respectively. The red region amounts to the secondary component
of GW190814. The brown and the purple regions show two different reports of
the pulsar J0030+0451 \protect\cite{Miller2019,Riley2019}.}
\label{massnjl1}
\end{figure}
As one can see from Fig. \ref{massnjl1}, there are different colored regions
in Fig. \ref{massnjl1}. The gray and orange areas are the $M-R$ constraints
from GW170817. The black boundary corresponds to PSR J0740+6620 \cite%
{Cromartie2019}. The pink area represents the secondary component of
GW190814 \cite{ZMiao2021}, and the green area shows the remnant mass of
GW190425 \cite{JSedaghat}. As one can see from this figure, we have not
drawn graphs for the values $\arrowvert C_{2}\arrowvert>0.14$, because Fig. %
\ref{tidal-alph=0.5-data1} shows for these values of $C_{2}$, the $\Lambda $
condition ($\Lambda _{1.4{M}_{\odot }}\lesssim 580$) is not satisfied. In
Fig. \ref{tidal-alph=0.5-data1}, the horizontal black line shows the $%
\Lambda =580$, and the vertical red line represents $M=1.4{M}_{\odot }$. As
one can see from this figure, the constraint $\Lambda _{1.4{M}_{\odot
}}\lesssim 580$, is satisfied for $\lvert C_{2}\lvert \lesssim 0.14$ which
corresponds to the $M_{TOV}\lesssim 2.13M_{\odot }$. Indeed, the maximum
mass in this case cannot be more than $2.13M_{\odot }$. More details for
structural properties of SQS for $B^{1/4}=117MeV$, $\alpha =0.5$, $%
C_{1}=10^{-5},$ and different values of $C_{2}$ are presented in Table. \ref%
{structural properties1}. Previously, in the Ref. \cite{ChengMingLi2020} the
value of $M_{TOV}$ for $B^{1/4}=117MeV$ and $\alpha =0.5$ has been
calculated in GR. The obvious difference between their result ($%
M_{TOV}=1.70M_{\odot }$) with the result obtained in this paper shows the
important effect of the massive gravity in the mass of the SQS. Table. \ref%
{structural properties1} demonstrates that by increasing the absolute value
of $C_{2}$, the values of $M_{TOV}$, $\Lambda _{1.4{M}_{\odot }}$, and the
compactness of the star are increased too. Additionally, The results
calculated for $R_{Sch}$ show that the obtained masses can not be black
holes.
\begin{table}[tbp]
\caption{Structural properties of SQS for $B^{1/4}=117MeV,\protect\alpha %
=0.5 $, $C_{1}=10^{-5},$ and different values of $C_{2}$.}
\label{structural properties1}\centering
\begin{tabular}{|c||c|c|c|c|}
\hline
$C_2 $ & $-10^{-3} $ & $-10^{-2}$ & $-0.1$ & $-0.14$ \\ \hline
$R(km)$ & $10.33$ & $10.36$ & $10.70$ & $11.02$ \\ \hline
$M_{TOV}({M}_\odot)$ & $1.76 $ & $1.78$ & $2.02$ & $2.13$ \\ \hline
$\Lambda_{1.4 {M}_{\odot }}$ & $446.76$ & $457.64$ & $548.92$ & $578.14$ \\
\hline
$\sigma (10^{-1})$ & $2.52$ & $2.54$ & $2.79$ & $2.86$ \\ \hline
$R_{Sch}(km)$ & $5.20$ & $5.22$ & $5.43$ & $5.53$ \\ \hline
\end{tabular}%
\end{table}

\begin{figure}[h]
\center{\includegraphics[width=8.5cm] {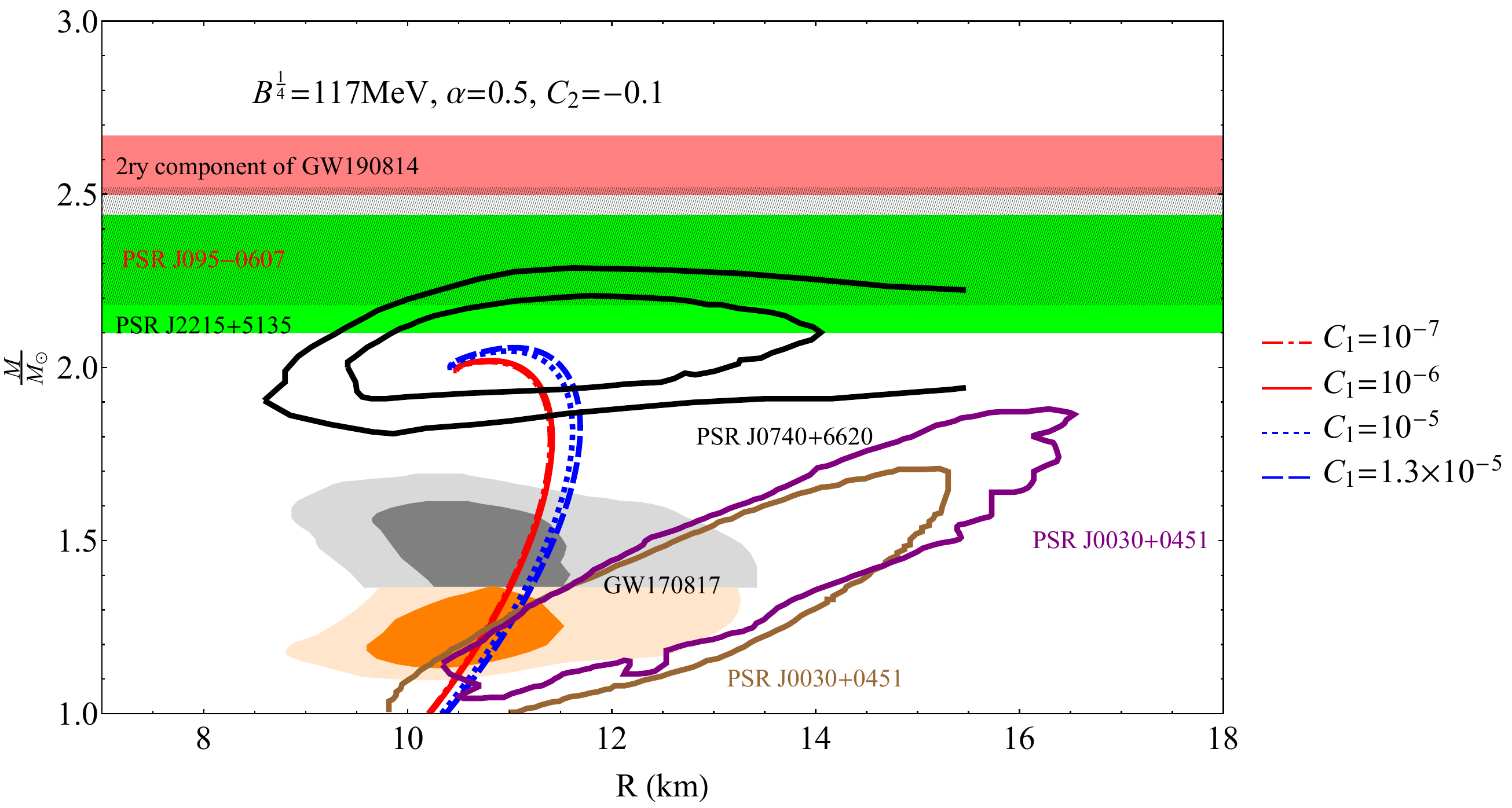}}
\caption{Mass-Radius diagram for $B^{1/4}=117MeV,\protect\alpha =0.5$, $%
C_{2}=-0.1$, and different values of $C_{1}$. The gray and orange regions
are the mass - radius constraints from the GW170817 event. The black region
shows pulsar J0740+6620, The green and black hatched regions represent
pulsars J2215+5135 \protect\cite{Linares2018} and PSR J095-0607,
respectively. The red region amounts to the secondary component of GW190814.
The brown and the purple regions show two different reports of the pulsar
J0030+0451 \protect\cite{Miller2019,Riley2019}.}
\label{massnjl2}
\end{figure}

Now, we turn to the next case by choosing $B^{1/4}=117MeV$, $\alpha =0.5$, $%
C_{2}=-0.1$ and different values of $C_{1}$. It is shown in Refs. \cite%
{Hendi2016,Behzad2017} that changing the $C_{1}$ parameter has a negligible
effect on the structural properties of the star. Therefore, in this section,
we show this feature and for other values of $\alpha $ and the bag constant,
we only change the value of $C_{2}$. Figure \ref{massnjl2} represents the $%
M-R$ relation for different values of $C_{1}$.
As this figure shows, changing the $C_{1}$ parameter has an insignificant
effect on the maximum mass of the star. We have not plotted the results for
values $C_{1}\gtrsim 1.3\times 10^{-5}$, because as the figure \ref%
{tidal-alph=0-data2} shows, the condition $\Lambda _{1.4{M}_{\odot
}}\lesssim 580$ is not established for these values of $C_{1}$. The
structural properties of SQS for $B^{1/4}=117MeV$, $\alpha =0.5$, $%
C_{2}=-0.1 $, and different values of $C_{1}$ are in Table. \ref{structural
properties2}.

On the other hand, our results in Table. \ref{structural properties2}
indicate that the obtained objects cannot be black holes because the radii
of these compact objects are bigger than the Schwarzschild radii. Besides,
the strength of gravity increases by increasing the value of $\lvert
C_{2}\lvert$.

\begin{figure}[h]
\center{\includegraphics[width=8.5cm] {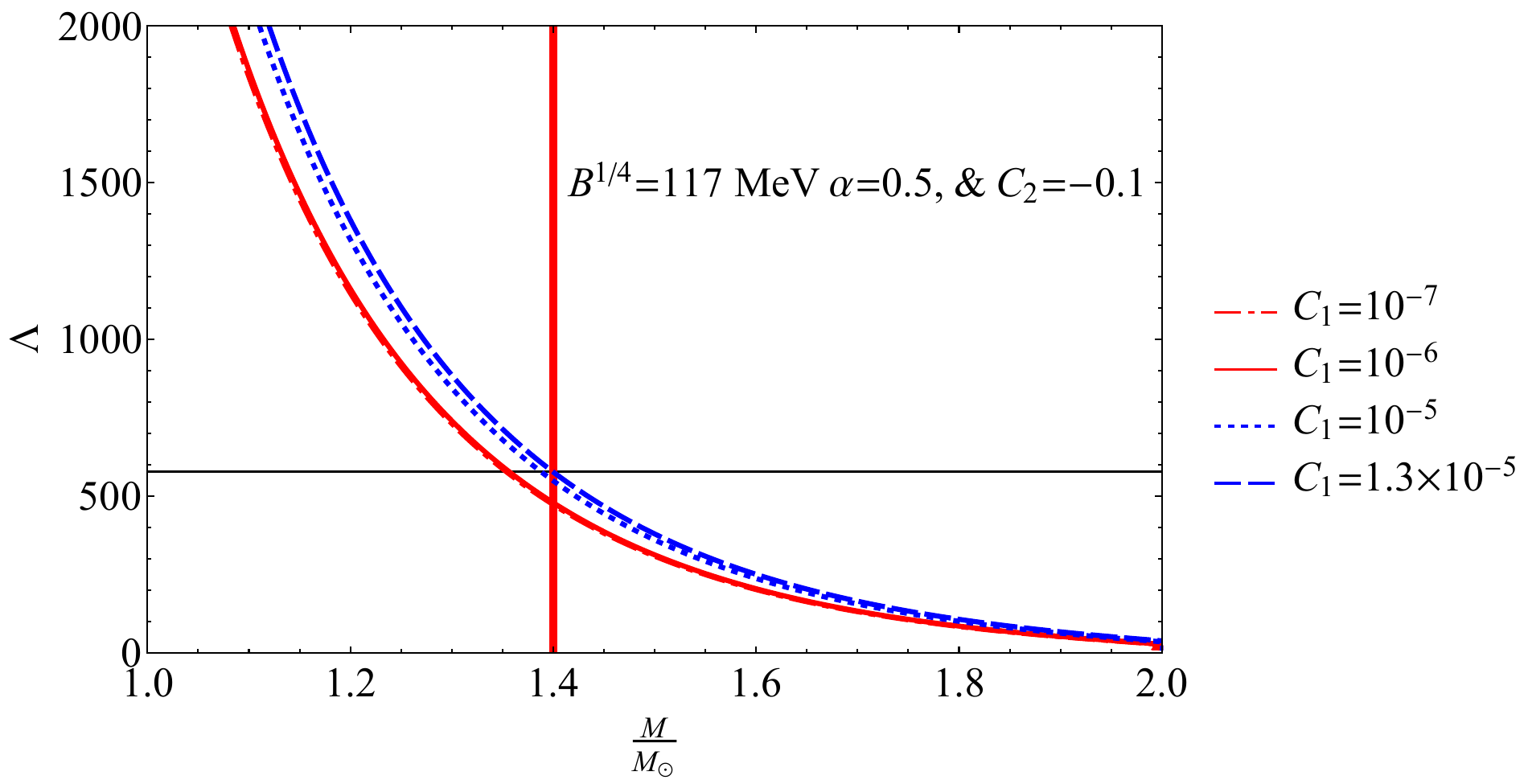}}
\caption{Dimensionless tidal deformability ($\Lambda$) versus mass of the
star for $B^{1/4}=117 MeV, \protect\alpha =0.5$, $C_2=-0.1$, and different
values of $C_1$. The horizontal black line and the vertical red line
correspond to $\Lambda=580$ and $\frac{M}{{M}_{\odot }}=1.4$, respectively.}
\label{tidal-alph=0-data2}
\end{figure}
\begin{table}[h]
\caption{Structural properties of SQS for $B^{1/4}=117 MeV, \protect\alpha %
=0.5$, $C_2=-0.1$, and different values of $C_1$.}
\label{structural properties2}\centering
\begin{tabular}{|c||c|c|c|c|}
\hline
$C_1 $ & $10^{-7} $ & $10^{-6}$ & $10^{-5}$ & $1.3\times10^{-5}$ \\ \hline
$R(km)$ & $10.61$ & $10.92$ & $10.95$ & $10.99$ \\ \hline
$M_{TOV}({M}_\odot)$ & $2.01 $ & $2.02$ & $2.05$ & $2.06$ \\ \hline
$\Lambda_{1.4 {M}_{\odot }}$ & $473.77$ & $480.89$ & $548.92$ & $578.25$ \\
\hline
$\sigma (10^{-1})$ & $2.80 $ & $2.74$ & $2.77$ & $2.77$ \\ \hline
$R_{Sch}(km)$ & $5.41 $ & $5.43$ & $5.52$ & $5.54$ \\ \hline
\end{tabular}%
\end{table}

\subsubsection{$\protect\alpha =0.8$}

\begin{figure}[h]
\center{\includegraphics[width=8.5cm] {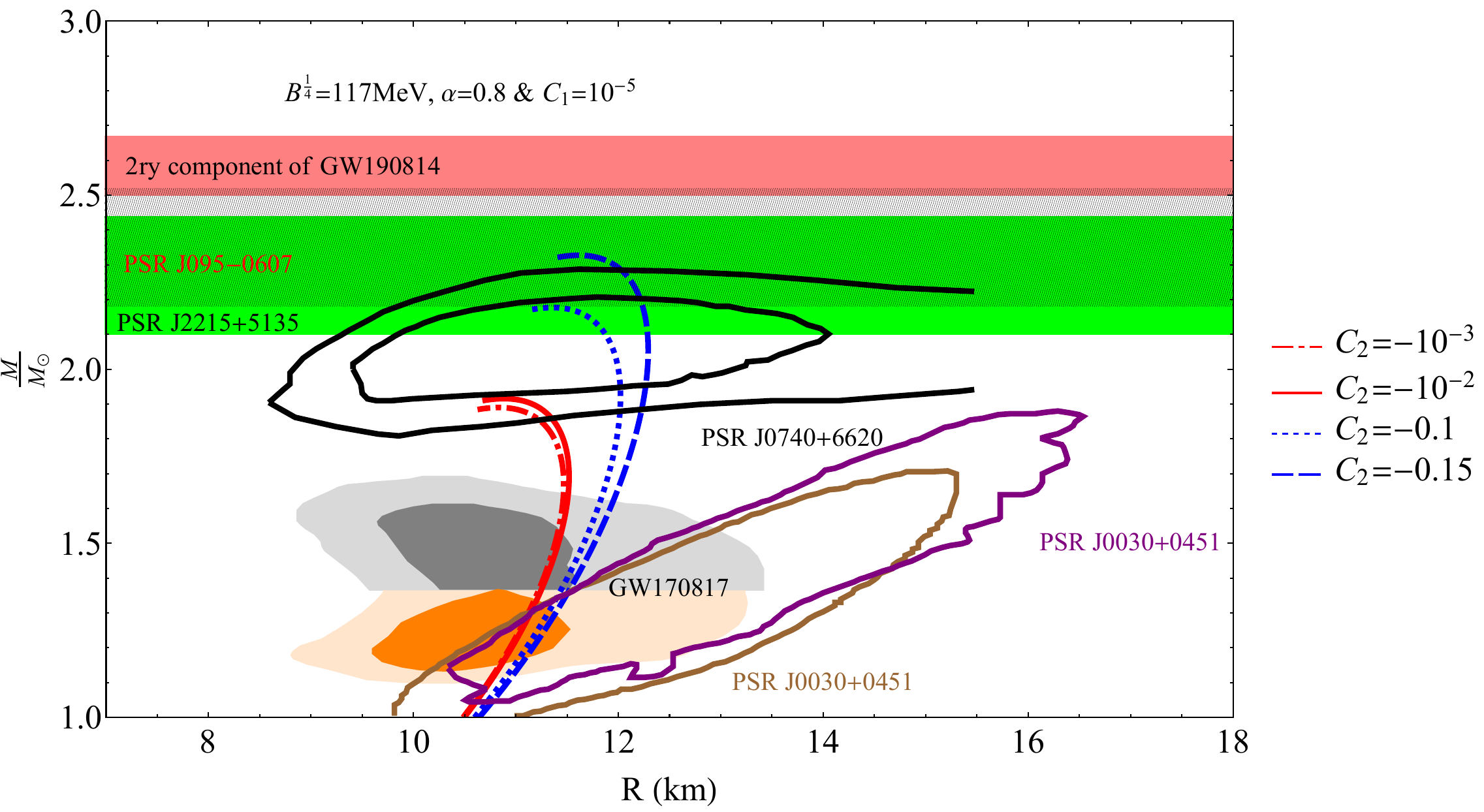}}
\caption{Mass-Radius diagram for $B^{1/4}=117MeV,\protect\alpha %
=0.8,C_{1}=10^{-5}$, and different values of $C_{2}$. The gray and orange
regions are the mass - radius constraints from the GW170817 event. The black
region shows pulsar J0740+6620, The green and black hatched regions
represent pulsars J2215+5135 \protect\cite{Linares2018} and PSR J095-0607,
respectively. The red region amounts to the secondary component of GW190814.
The brown and the purple regions show two different reports of the pulsar
J0030+0451 \protect\cite{Miller2019,Riley2019}.}
\label{mass-njl1-0.8}
\end{figure}

\begin{figure}[h]
\center{\includegraphics[width=8.5cm] {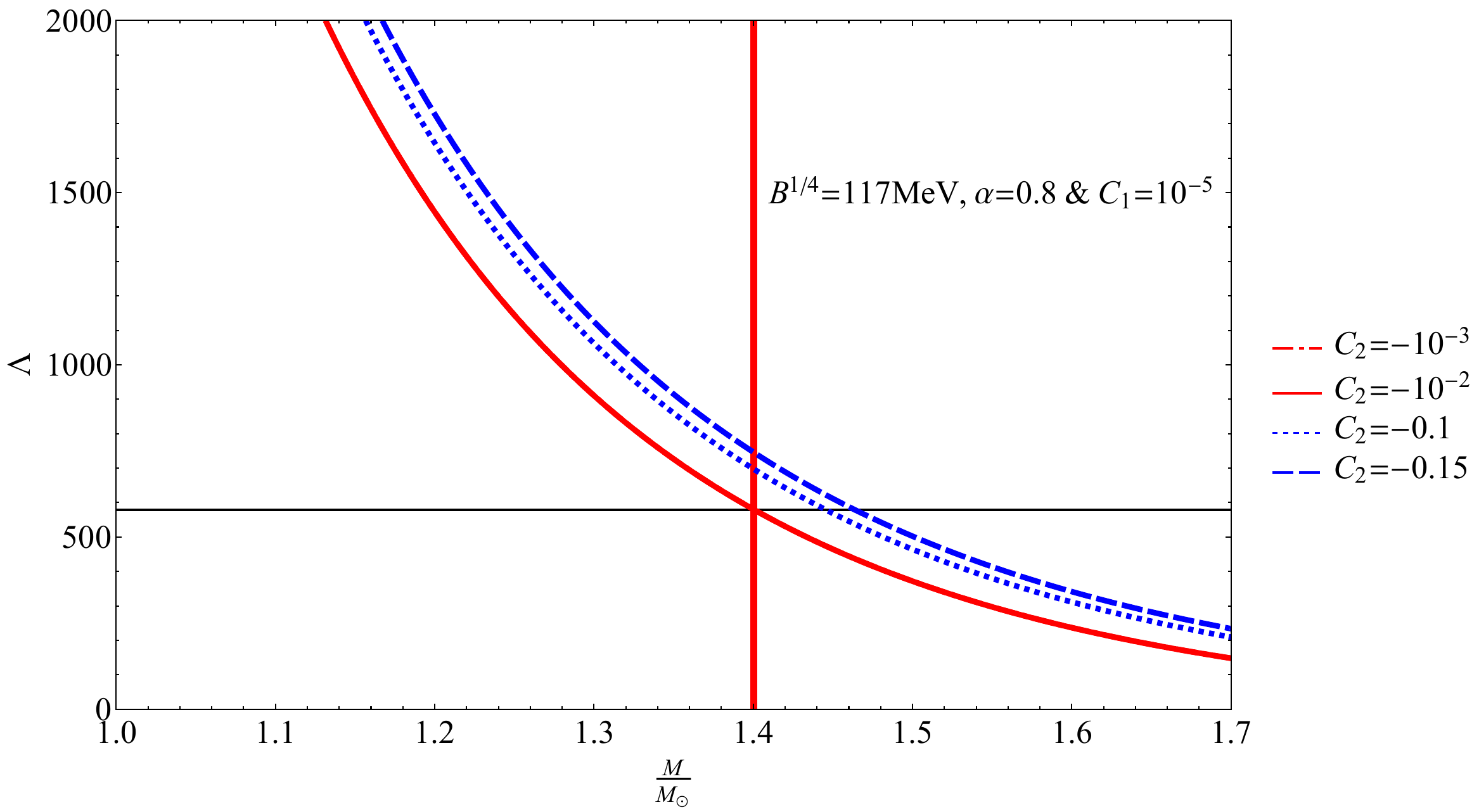}}
\caption{Dimensionless tidal deformability versus mass of the star for $%
B^{1/4}=117MeV,\protect\alpha =0.8,\hspace{1mm}C_{1}=10^{-5},$ and different
values of $C_{2}$. The horizontal black line and the vertical red line
correspond to $\Lambda =580$ and $\frac{M}{{M}_{\odot }}=1.4$, respectively.}
\label{data-of-tidal1-0.8}
\end{figure}

Here, we increase the value of $\alpha $ and investigate its effect on the
structure properties of the SQS in dRGT-like massive gravity. The figures. %
\ref{mass-njl1-0.8} and \ref{data-of-tidal1-0.8} show the $M-R$ and $\Lambda
-M$ diagrams for $\alpha =0.8$,$\hspace{1mm}C_{1}=10^{-5}$, and different
values of $C_{2}$, respectively.

Comparison of figures \ref{mass-njl1-0.8} and \ref{massnjl1} shows that the
maximum mass of the star increases with the increase of $\alpha $ value.
This result is obvious because we showed in Fig. \ref{eos-njl} that the
increase of $\alpha $ makes the EOS stiffer. However, Fig. \ref%
{data-of-tidal1-0.8} shows that this feature increases $\Lambda $ and the
constraint $\Lambda _{1.4{M}_{\odot }}\lesssim 580$ is lost for $\arrowvert %
C_{2}\arrowvert\gtrsim 10^{-2}$ which corresponds to the $M_{TOV}\lesssim
2.33M_{\odot }$. Indeed, the maximum mass in this case cannot be more than $%
2.33M_{\odot }$. The structural properties of SQS for $B^{1/4}=117MeV$, $%
\alpha =0.8$, $C_{1}=10^{-5}$, and different values of $C_{2}$ are presented
in Table. \ref{structural properties3}.

There are the same behaviors for the compactness and the modified
Schwarzschild radius. They increase by increasing the value of $\lvert
C_{2}\lvert$. Also, the radii of these compact objects are more than the
modified Schwarzschild radii (i.e., $R>R_{Sch}$).
\begin{table}[h]
\caption{Structural properties of SQS for $B^{1/4}=117MeV, \protect\alpha %
=0.8$, $C_{1}=10^{-5}$, and different values of $C_{2}$.}
\label{structural properties3}\centering
\begin{tabular}{|c||c|c|c|c|}
\hline
$C_2 $ & $-10^{-3} $ & $-10^{-2}$ & $-0.1$ & $-0.15$ \\ \hline
$R(km)$ & $10.80$ & $10.92$ & $11.33$ & $11.61$ \\ \hline
$M_{TOV}({M}_\odot)$ & $1.89 $ & $1.92$ & $2.18$ & $2.33$ \\ \hline
$\Lambda_{1.4 {M}_{\odot }}$ & $578.35$ & $578.69$ & $700.07$ & $745.67$ \\
\hline
$\sigma (10^{-1})$ & $2.59$ & $2.60$ & $2.85$ & $2.97$ \\ \hline
$R_{Sch}(km)$ & $5.59$ & $5.63$ & $5.87$ & $6.00$ \\ \hline
\end{tabular}%
\end{table}

\subsubsection{$\protect\alpha =0.94$}

\begin{figure}[h]
\center{\includegraphics[width=8.5cm] {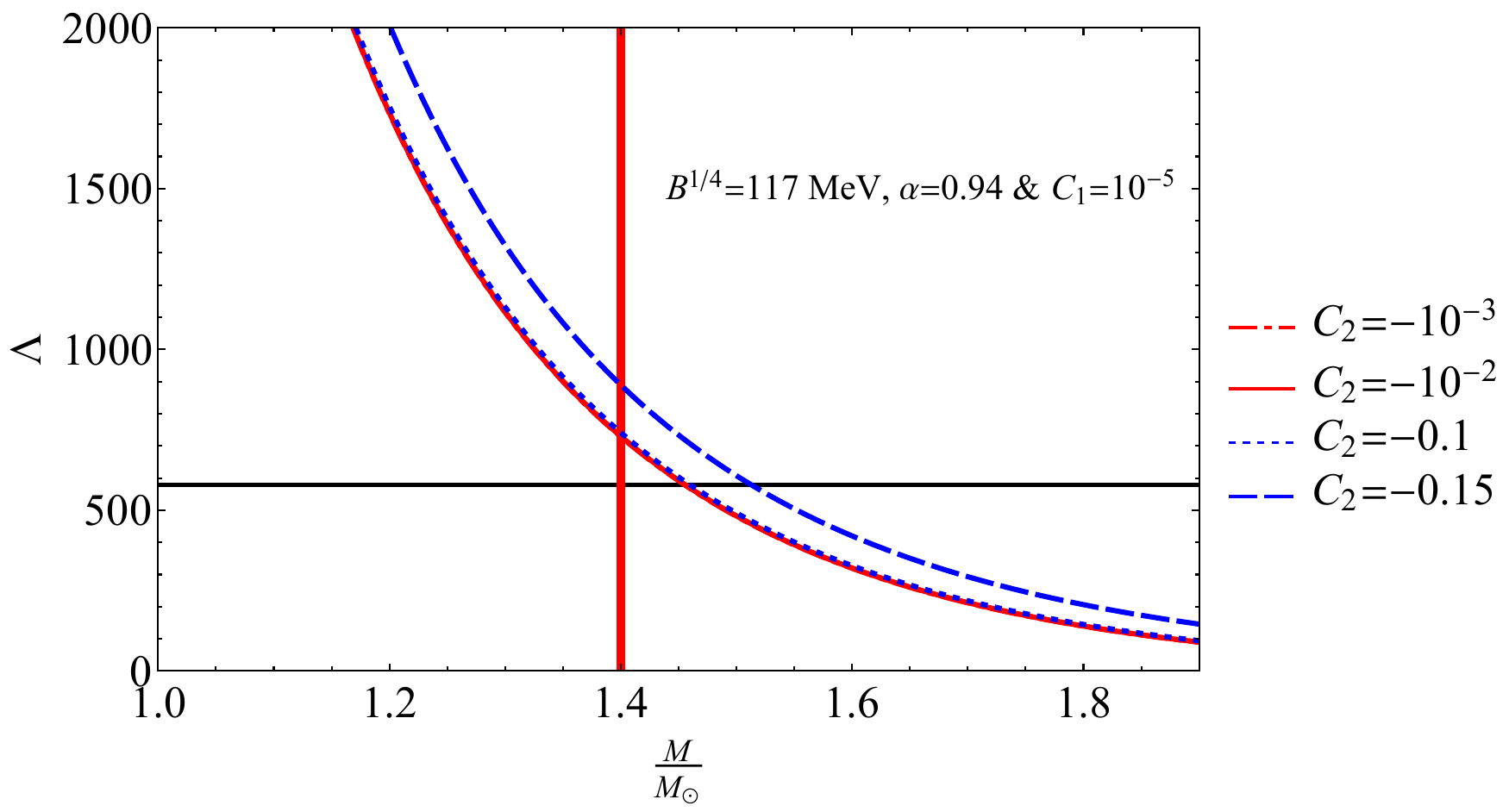}}
\caption{Dimensionless tidal deformability versus mass of the star for $%
B^{1/4}=117MeV$, $\protect\alpha=0.94$, $C_{1}=10^{-5}$, and different
values of $C_{2}$. The horizontal black line and the vertical red line
correspond to $\Lambda=580$ and $\frac{M}{{M}_{\odot }}=1.4$, respectively.}
\label{dataoftidal5}
\end{figure}

For further investigation, we check the effect of massive gravity on the
structural properties of SQS for $\alpha =0.94$. We do not consider values
of $\alpha$ greater than $0.94$. Because, as mentioned in section \ref{mnjl}%
, the stability condition of SQM will be lost for $\alpha\gtrsim 0.94$ \cite%
{ChengMingLi2020}. BY setting $B^{1/4}=117 MeV$, $\alpha =0.94$, $%
C_{1}=10^{-5}$, and different values of $C_{2}$, we obtain the Fig. \ref%
{massnjl5}, and \ref{dataoftidal5} as the $M-R$ and $\Lambda -M$ diagrams,
respectively. Here, we also see that with the increase of $\alpha $, the
maximum gravitational mass increases, but as the EOS becomes stiffer, the $%
\Lambda$ constraint is no longer satisfied. Table. \ref{structural
properties5} shows the structural properties of SQS for $B^{1/4}=117 MeV$, $%
\alpha =0.94$, $C_{1}=10^{-5}$, and different values of $C_{2}$. Although
the values obtained for the $M_{TOV}$ for $\alpha=0.94$ are higher than
those with $\alpha=0.5$ and $\alpha=0.8$, Table. \ref{structural properties5}
demonstrates that none of the results satisfy the constraint $%
\Lambda_{1.44M_{\odot }}\lesssim580$.

\begin{figure}[h]
\center{\includegraphics[width=8.5cm] {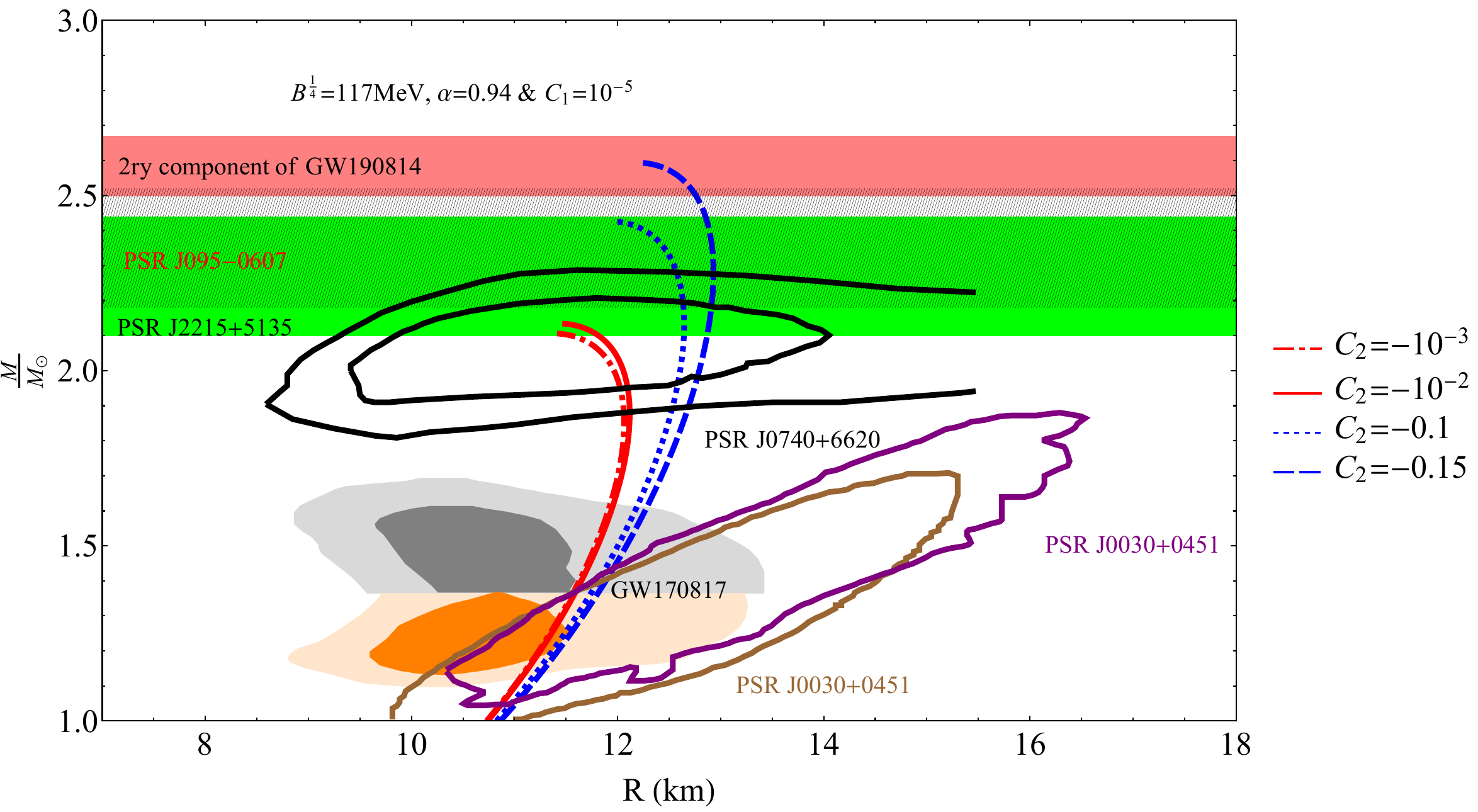}}
\caption{Mass-Radius diagram for $B^{1/4}=117MeV$, $\protect\alpha =0.94$, $%
C_{1}=10^{-5}$, and different values of $C_{2}$. The gray and orange regions
are the mass - radius constraints from the GW170817 event. The black region
shows pulsar J0740+6620, The green and black hatched regions represent
pulsars J2215+5135 \protect\cite{Linares2018} and PSR J095-0607,
respectively. The red region amounts to the secondary component of GW190814.
The brown and the purple regions show two different reports of the pulsar
J0030+0451 \protect\cite{Miller2019,Riley2019}.}
\label{massnjl5}
\end{figure}
\begin{table}[tbp]
\caption{Structural properties of SQS for $B^{1/4}=117MeV, \protect\alpha %
=0.94$, $C_{1}=10^{-5}$, and different values of $C_{2}$.}
\label{structural properties5}\centering
\begin{tabular}{|c||c|c|c|c|}
\hline
$C_2 $ & $-10^{-3} $ & $-10^{-2}$ & $-0.1$ & $-0.15$ \\ \hline
$R(km)$ & $11.41$ & $11.46$ & $12.06$ & $12.23$ \\ \hline
$M_{TOV}({M}_\odot)$ & $2.11$ & $2.13$ & $2.43$ & $2.60$ \\ \hline
$\Lambda_{1.4 {M}_{\odot }}$ & $728.30$ & $733.20$ & $747.51$ & $887.92$ \\
\hline
$\sigma (10^{-1})$ & $2.74$ & $2.75$ & $2.98$ & $3.09$ \\ \hline
$R_{Sch}(km)$ & $6.24$ & $6.30$ & $6.54$ & $6.70$ \\ \hline
\end{tabular}%
\end{table}

As an important result of our calculation, we could not describe the mass
gap region and observational objects such as PSRJ0740+6620, the secondary
component of GW190814, and the remnant mass of GW190425 by SQS when we
considered the mentioned EOSs of SQM with $B^{1/4}=117MeV$ and different
values of $\alpha$. In other words, the maximum mass of SQS with the
mentioned EOSs in dRGT-like massive gravity is in the range $%
M_{TOV}<2.5M_{\odot }$.

To obtain massive SQSs that satisfy the $\Lambda$ constraint, we make the
EOS soft. Because the value of $\Lambda$ depends on both the EOS and the
gravity used, simultaneously. Therefore, in the next section, we change the
value of the $B$ constant (as we expected) and investigate the results for
thermodynamic properties of SQM and structural properties of SQS in
dRGT-like massive gravity.

\section{Equation of States of SQM for $B^{1/4}=130MeV$}

\begin{figure}[h]
\center{\includegraphics[width=8.5cm] {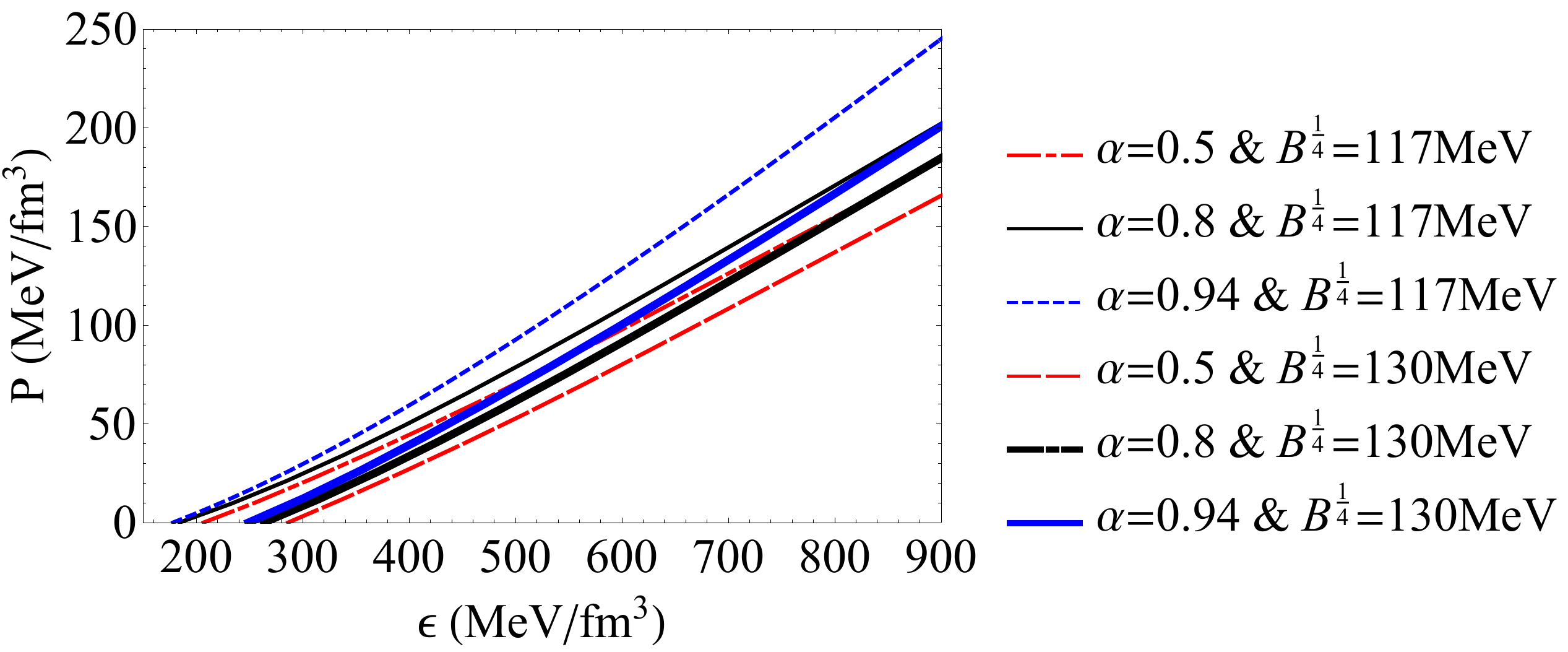}}
\caption{Comparison between EOSs for different values of $B$.}
\label{eos-comparing}
\end{figure}

In the previous section, we saw that by increasing the value of $\alpha $,
the EOS becomes stiffer, and subsequently the mass of the star increases,
but the constraint $\Lambda _{1.44M_{\odot }}\lesssim580$ is not satisfied
for the SQSs fall within the mass gap region and consequently, a small range
of the $C_{2}$ parameter is reachable. In this section, we increase the
value of the bag constant, $B$, to make the EOS softer, and the condition of
$\Lambda $ to be satisfied in a larger range of $C_{2}$ parameter. Figure %
\ref{eos-comparing} shows a comparison between two EOSs obtained from two
different bag constants ($B^{1/4}=117MeV\ $and$\ B^{1/4}=130MeV$). From this
figure, it can be seen that the EOSs with the same value of $\alpha $ have
the same slope and consequently have the same speed of sound. But the
diagrams with $B^{1/4}=130MeV$ lie lower than those with $B^{1/4}=117MeV$.
It shows that the EOSs with $B^{1/4}=130MeV$, are softer than those with $%
B^{1/4}=130MeV$. Although this result is obvious, we will indicate that it
has a significant effect on the structural properties of SQS in massive
gravity. This makes it possible to obtain quark stars that fall within the
mass gap region and respect to the constraint $\Lambda _{1.4M_{\odot
}}\lesssim580$, simultaneously. Now, we set $B^{1/4}=130MeV$ and investigate
its effect on the structural properties of SQS in massive gravity. In this
situation, we consider another constraint too. We assume that the value of $%
\Lambda$ for $M_{TOV}$ should not be zero. Because otherwise the obtained
mass is a black hole \cite{Chia2021}.

\begin{figure}[h]
\center{\includegraphics[width=8.5cm] {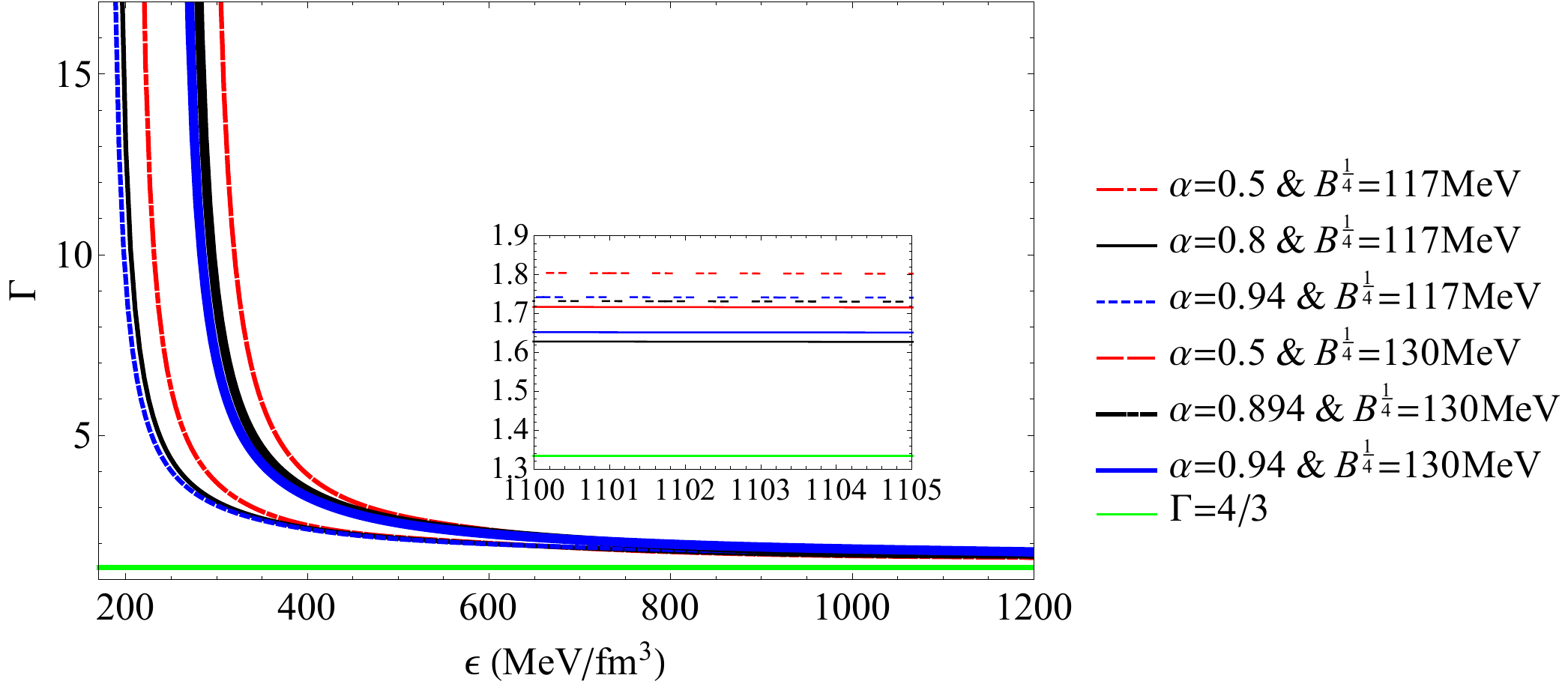}}
\caption{Adiabatic index ($\Gamma$) for different values $\protect\alpha$
and $B$.}
\label{adia-comparing}
\end{figure}

Figure. \ref{adia-comparing} represents the adiabatic index for different
values of $\alpha$ and $B$. As this figure shows, for any choice of $\alpha$%
, dynamical stability is increased by increasing $B$.

\subsection{Structural Properties of SQS in Massive Gravity for $%
B^{1/4}=130MeV$}

In the following, we study the structural properties of SQS in the presence
of dRGT-like massive gravity for $B^{1/4}=130MeV$, and three different
values of $\alpha=0.5$, $\alpha=0.8$, and $\alpha=0.94$.

\subsubsection{$\protect\alpha =0.5$}

\begin{figure}[h]
\center{\includegraphics[width=8.5cm] {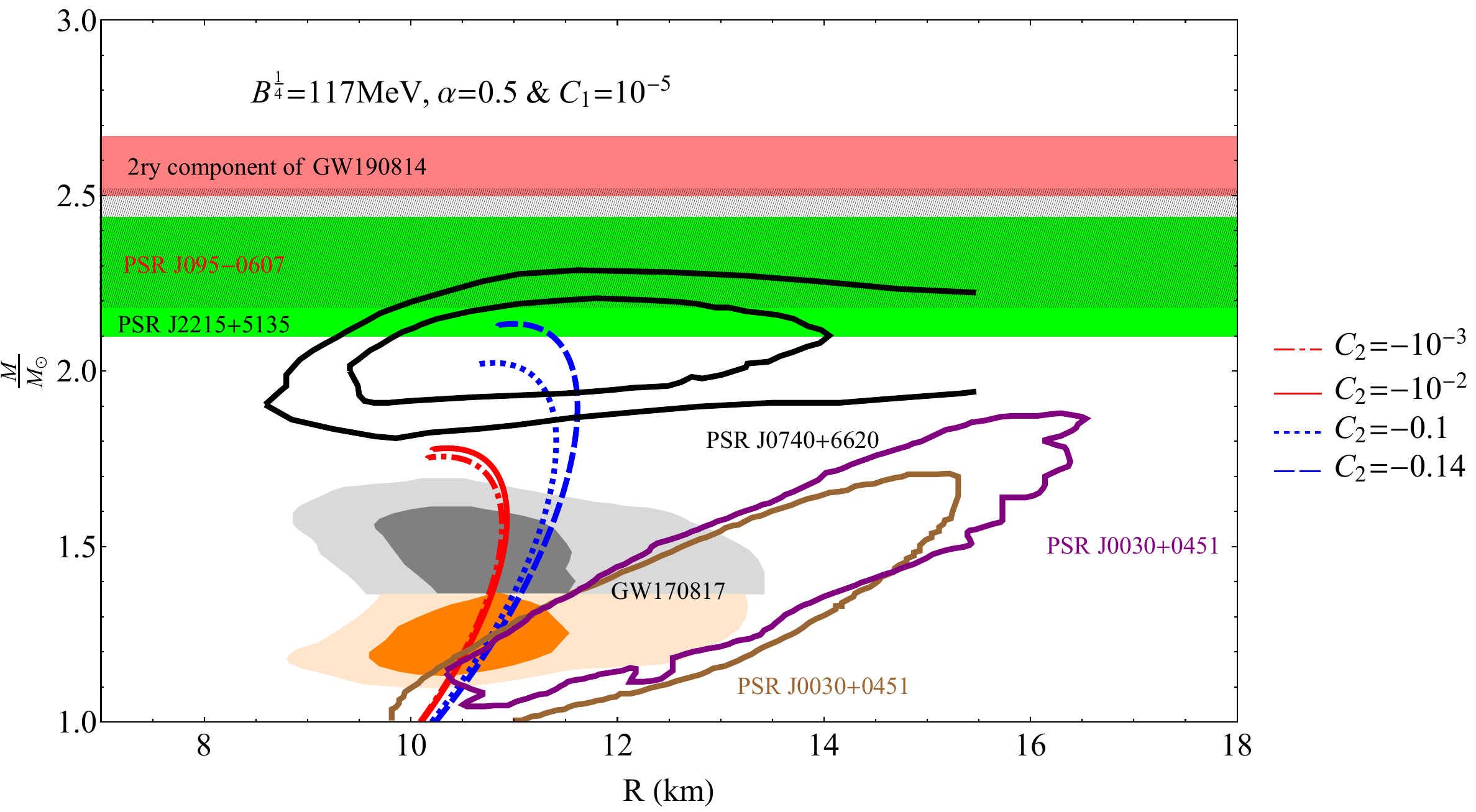}}
\caption{Mass-Radius diagram for $B^{1/4}=130MeV$, $\protect\alpha=0.5$, $%
C_{1}=10^{-5}$, and different values of $C_2$. The gray and orange regions
are the mass - radius constraints from the GW170817 event. The black region
shows pulsar J0740+6620, The green and black hatched regions represent
pulsars J2215+5135 \protect\cite{Linares2018} and PSR J095-0607,
respectively. The red region amounts to the secondary component of GW190814.
The brown and the purple regions show two different reports of the pulsar
J0030+0451 \protect\cite{Miller2019,Riley2019}. The light blue band denotes
the remnant mass of GW190425.}
\label{massnjl7}
\end{figure}

\begin{figure}[h]
\center{\includegraphics[width=8.5cm] {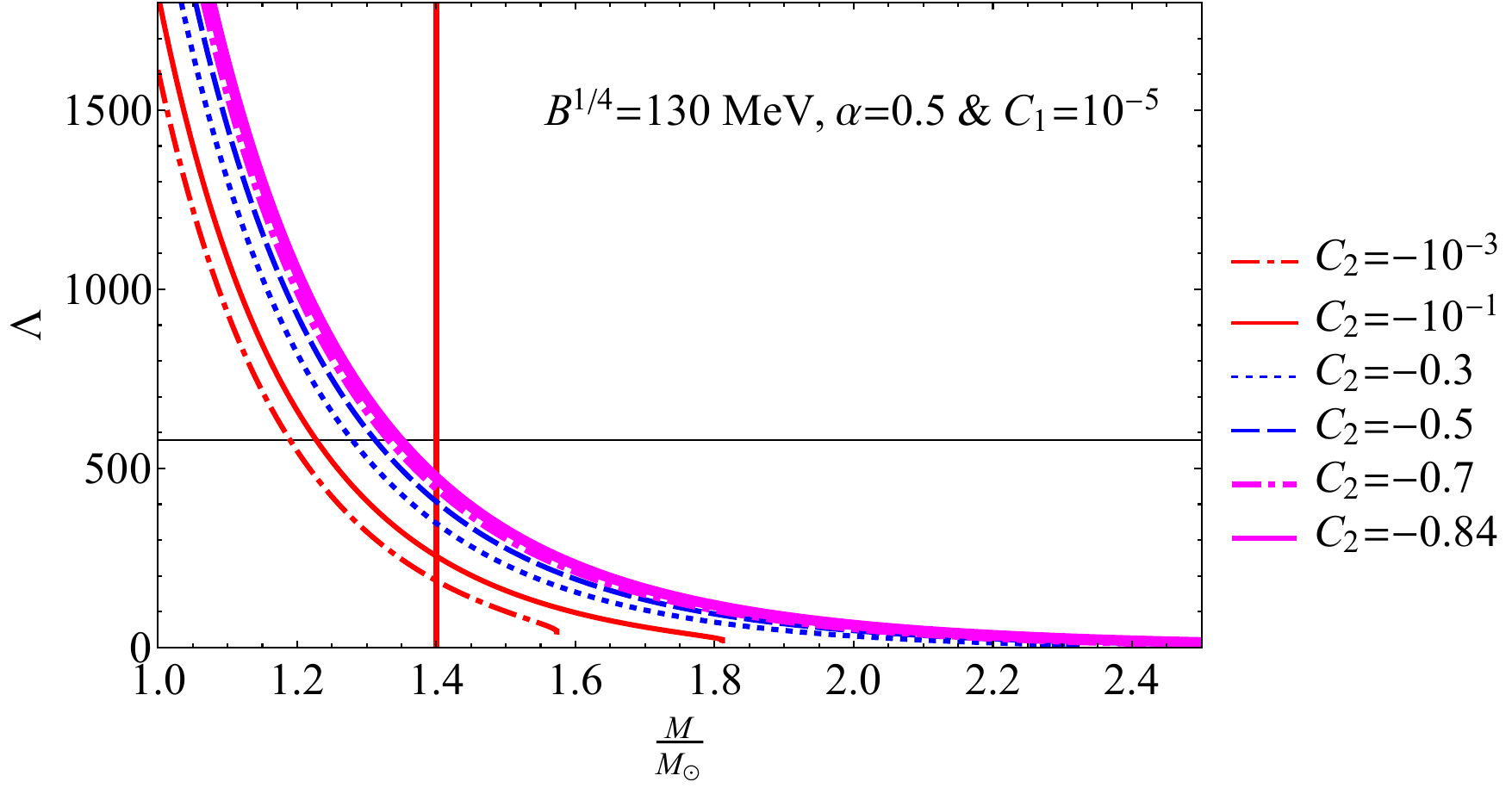}}
\caption{Dimensionless tidal deformability ($\Lambda$) versus mass of the
star for $B^{1/4}=130MeV$, $\protect\alpha =0.5$, $C_{1}=10^{-5}$, and
different values of $C_2$. The horizontal black line and the vertical red
line correspond to $\Lambda=580$ and $\frac{M}{{M}_{\odot }}=1.4$,
respectively.}
\label{dataoftidal7}
\end{figure}

Here, we derive the structural properties of SQS in massive gravity for $%
\alpha=0.5$ and $B^{1/4}=130 MeV$. Similar to the previous section, we first
set a fixed value for $C_1$ and investigate $M-R$ and $\Lambda-M$ diagrams
for different values of $C_2$. 
Figures \ref{massnjl7} and \ref{dataoftidal7} show $M-R$ and $\Lambda -M$
diagrams, respectively. We can see from the fig. \ref{dataoftidal7} that all
the diagrams satisfy the constraint $\Lambda _{1.4M_{\odot }}\lesssim580$.
The results show that by softening the EOS, the masses obtained in massive
gravity not only satisfy the $\Lambda_{1.4 {M}_{\odot }}\lesssim580$
constraint, but also some of them are located in the mass gap region. As we
can see from these figures, we did not use the values $\arrowvert C_{2}%
\arrowvert\gtrsim 0.84$, because they led to zero value for $%
\Lambda_{M_{TOV}}$. The results for structural properties are given in
Table. \ref{structural properties7}.

As an important result, our findings indicate that the maximum mass of SQS
can be in the range $M_{TOV}\lesssim 4M_{\odot }$. Indeed, by using the MNJL
model with $B^{1/4}=130 MeV$, and $\alpha=0.5$, SQSs fall within the mass
gap region. 
\begin{table*}[tbp]
\caption{Structural properties of SQS for $B^{1/4}=130MeV$, $\protect\alpha %
=0.5$, $C_{1}=10^{-5}$, and different values of $C_2$. The results falling
in mass gap have been separated in different column.}
\label{structural properties7}\centering
\begin{tabular}{|c||c|c|c|c|c|c|c|c|c|}
\hline
\multicolumn{4}{c}{none mass gap} &  & \multicolumn{5}{c}{mass gap} \\ \hline
$C_2 $ & $-10^{-3}$ & $-0.1$ & $-0.2$ & $-0.3$ & $-0.4$ & $-0.5$ & $-0.6$ & $%
-0.7$ & $-0.84$ \\ \hline
$R(km)$ & $9.04 $ & $9.46$ & $9.85$ & $10.24$ & $10.66$ & $11.06$ & $11.40$
& \hspace{0.1cm} $11.75$ & $12.34$ \\ \hline
$M_{TOV}({M}_\odot)$ & $1.56 $ & $1.80$ & $2.05$ & $2.31$ & $2.57$ & $2.86$
& $3.15$ & \hspace{0.1cm} $3.45$ & $3.91$ \\ \hline
$\Lambda_{1.4 {M}_{\odot }}$ & $185.80$ & $255.06$ & $307.00$ & $348.34$ & $%
379.38$ & $403.35$ & $431.03$ & $449.16$ & \hspace{0.1cm} $480.18$ \\ \hline
$\Lambda_{M_{TOV}}$ & $59.11$ & $26.20$ & $12.74$ & $6.20$ & $3.30$ & $1.42$
& $0.64$ & $0.26$ & \hspace{0.1cm} $0.02$ \\ \hline
$\sigma (10^{-1})$ & $2.55 $ & $2.82$ & $3.08$ & $3.34$ & $3.57$ & $3.83$ & $%
4.09$ & \hspace{0.1cm} $4.34$ & $4.69$ \\ \hline
$R_{Sch}(km) $ & $4.61 $ & $4.84$ & $5.06$ & $5.26$ & $5.43$ & $5.64$ & $%
5.83 $ & \hspace{0.1cm} $6.00$ & $6.29$ \\ \hline
\end{tabular}%
\end{table*}

\subsubsection{$\protect\alpha =0.8$}

\begin{figure}[h]
\center{\includegraphics[width=8.5cm] {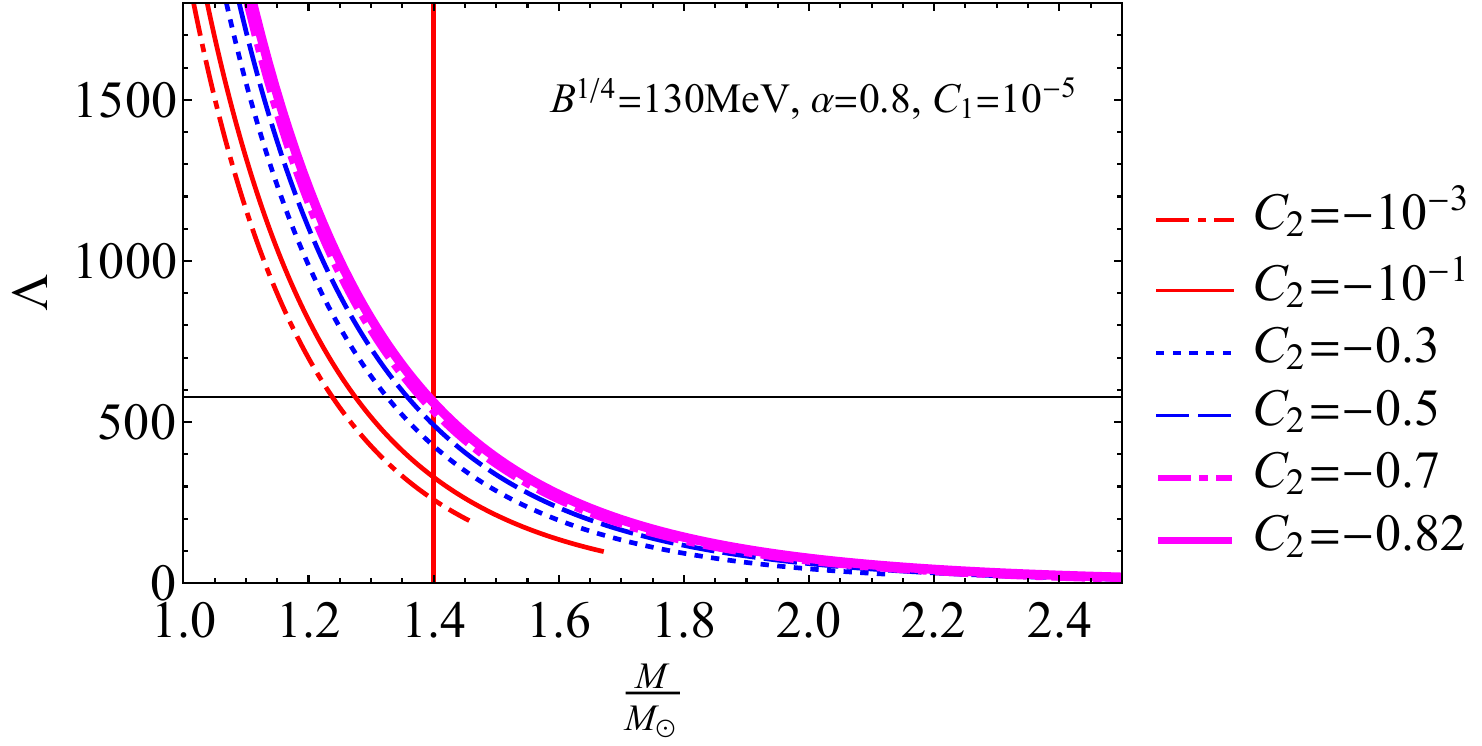}}
\caption{Dimensionless tidal deformability ($\Lambda$) versus mass of the
star for $B^{1/4}=130MeV$, $\protect\alpha =0.8$, $C_{1}=10^{-5}$, and
different values of $C_2$. The horizontal black line and the vertical red
line correspond to $\Lambda=580$ and $\frac{M}{{M}_{\odot }}=1.4$,
respectively.}
\label{dataoftidal8}
\end{figure}

As a further investigation, we change the value of $\alpha$. Figures. \ref%
{massnjl8}, and \ref{dataoftidal8} are $M-R$ and $\Lambda-M$ diagrams for $%
B^{1/4}=130MeV$, $\alpha=0.8$, $C_{1}=10^{-5}$ and different values of $C_2$%
, respectively. We can see that despite the increase of $\alpha$, which
leads to the stiffer EOS, the constraint $\Lambda_{1.4 {M}_{\odot
}}\lesssim580$ is still maintained. Here, the obtained masses lie in a
larger area of the mass gap region compared to the previous case ($%
B^{1/4}=130 MeV$ and $\alpha=0.5$). Moreover, we checked the $\Lambda $ of $%
M_{TOV}$ for different values of $C_2$. We observed that for $\arrowvert C_2%
\arrowvert\gtrsim0.84$ the values of $\Lambda _{M_{TOV}}$ became zero. The
results for structural properties are given in Table. \ref{structural
properties8}.

On the other hand, our calculation for the compactness and the modified
Schwarzschild radius show that these compact objects cannot be black holes
because their radii are bigger than the Schwarzschild radii, i.e., $%
R>R_{Sch} $ (see Table. \ref{structural properties8}, for more details).
Also, the compactness increases by increasing the value of $\lvert
C_{2}\lvert$. In other words, the strength of gravity augments when the
value of $\lvert C_{2}\lvert$ increases (see Table. \ref{structural
properties8}, for more details).
\begin{figure}[h]
\center{\includegraphics[width=8.5cm] {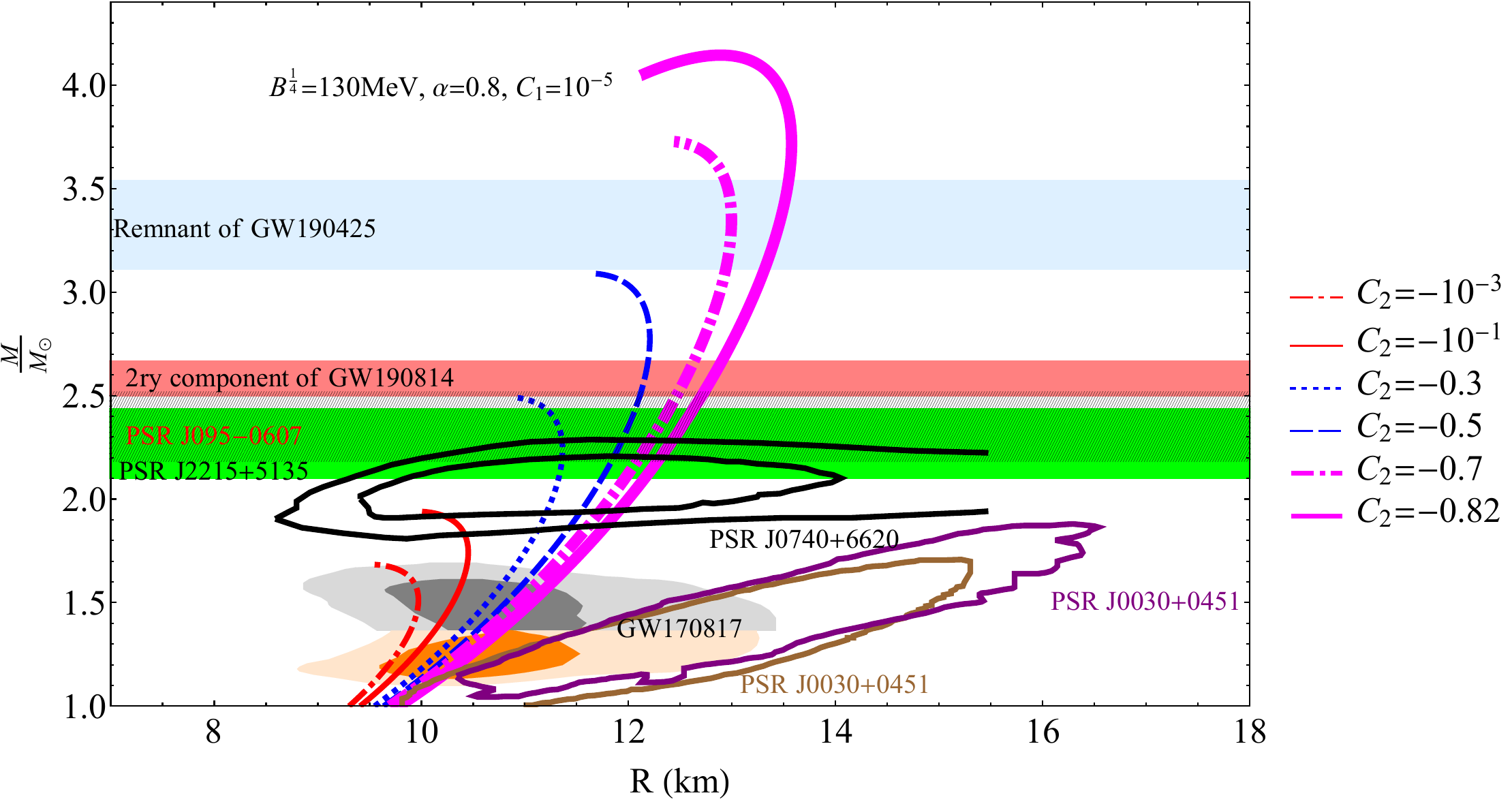}}
\caption{Mass-Radius diagram for $B^{1/4}=130MeV$, $\protect\alpha =0.8$, $%
C_{1}=10^{-5}$, and different values of $C_2$. The gray and orange regions
are the mass - radius constraints from the GW170817 event. The black region
shows pulsar J0740+6620, The green and black hatched regions represent
pulsars J2215+5135 \protect\cite{Linares2018} and PSR J095-0607,
respectively. The red region amounts to the secondary component of GW190814.
The brown and the purple regions show two different reports of the pulsar
J0030+0451 \protect\cite{Miller2019,Riley2019}. The light blue band denotes
the remnant mass of GW190425.}
\label{massnjl8}
\end{figure}
\begin{table*}[tbp]
\caption{Structural properties of SQS for $B^{1/4}=130MeV$, $\protect\alpha %
=0.8$, $C_{1}=10^{-5}$, and different values of $C_2$. The results falling
in mass gap have been separated in different column.}
\label{structural properties8}\centering
\begin{tabular}{|c||c|c|c|c|c|c|c|c|c|}
\hline
\multicolumn{3}{c}{none mass gap} &  & \multicolumn{6}{c}{mass gap} \\ \hline
$C_2 $ & $-10^{-3} $ & $-0.1$ & $-0.2$ & $-0.3$ & $-0.4$ & $-0.5$ & $-0.6$ &
$-0.7$ & $-0.82$ \\ \hline
$R(km)$ & $9.46 $ & $9.94$ & $10.40$ & $10.78$ & $11.15$ & $11.60$ & $11.97 $
& \hspace{0.1cm} $12.36$ & $12.80$ \\ \hline
$M_{TOV}({M}_\odot)$ & $1.68 $ & $1.94$ & $2.21$ & $2.49$ & $2.79$ & $3.09$
& $3.40$ & \hspace{0.1cm} $3.73$ & $4.14$ \\ \hline
$\Lambda_{1.4 {M}_{\odot }} $ & $257.98$ & $329.76$ & $383.76$ & $427.33$ & $%
461.80$ & $487.14$ & $515.31$ & \hspace{0.1cm} $536.06$ & $559.21$ \\ \hline
$\Lambda_{M_{TOV}} $ & $47.04$ & $20.43$ & $9.64$ & $2.70$ & $1.95$ & $0.92$
& $0.39$ & \hspace{0.1cm} $0.14$ & $0.01$ \\ \hline
$\sigma (10^{-1})$ & $2.63 $ & $2.89$ & $3.14$ & $3.42$ & $3.70$ & $3.94$ & $%
4.20$ & \hspace{0.1cm} $4.47$ & $4.79$ \\ \hline
$R_{Sch}(km) $ & $4.97$ & $5.22$ & $5.45$ & $5.67$ & $5.90$ & $6.10$ & $6.29$
& $6.49$ & $6.73$ \\ \hline
\end{tabular}%
\end{table*}

It is notable that by considering the MNJL model with $B^{1/4}=130 MeV$, and
$\alpha=0.8$, the maximum mass of SQSs can be in the range $%
1.68M_{\odot}\lesssim M_{TOV}\lesssim 4.2M_{\odot}$, which fall within the
mass gap region.

\subsubsection{$\protect\alpha =0.94$}

\begin{figure}[h]
\center{\includegraphics[width=8.5cm] {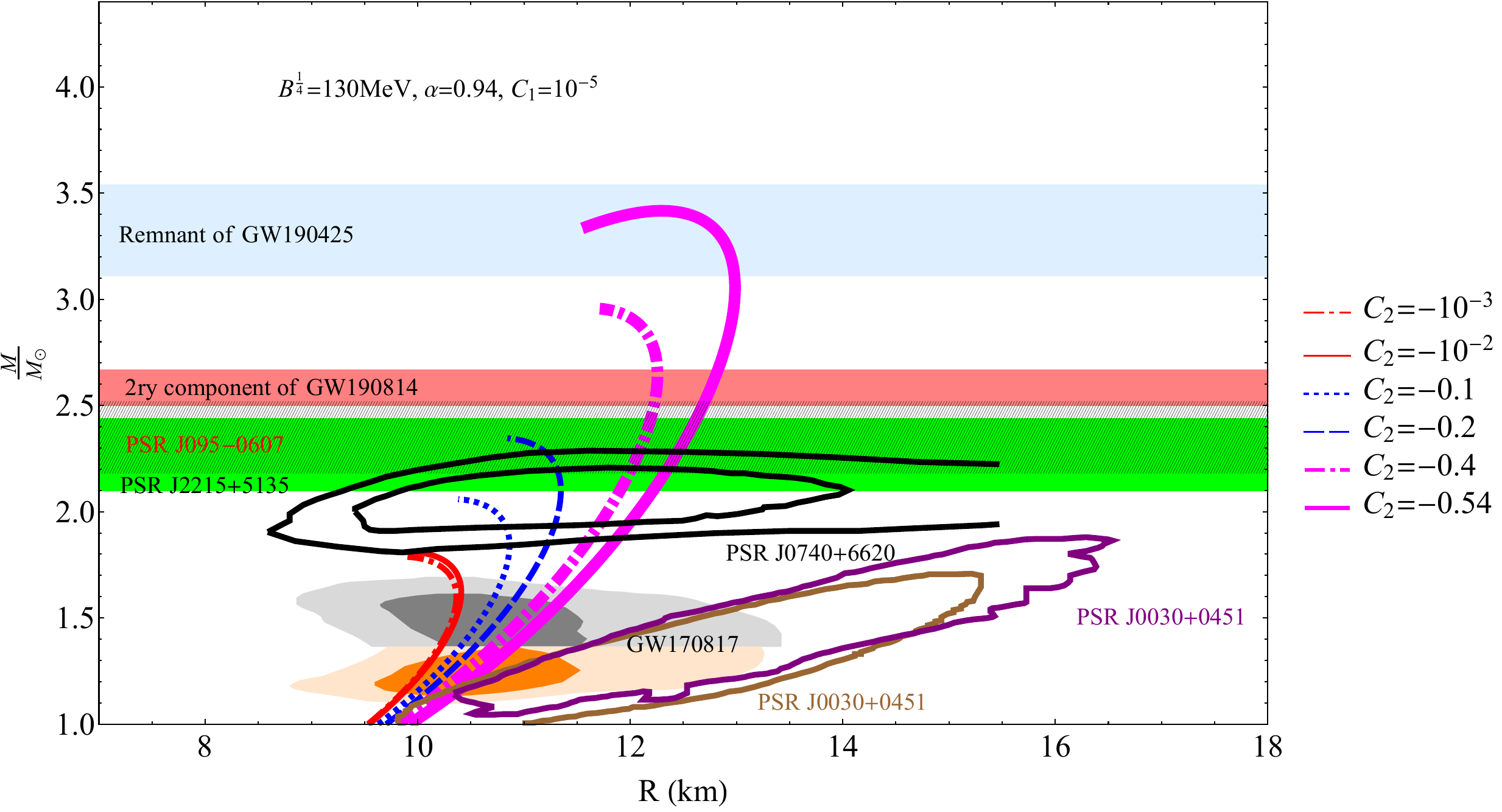}}
\caption{Mass-Radius diagram for $B^{1/4}=130MeV$, $\protect\alpha =0.94$, $%
C_{1}=10^{-5}$, and different values of $C_2$. The gray and orange regions
are the mass - radius constraints from the GW170817 event. The black region
shows pulsar J0740+6620, The green and black hatched regions represent
pulsars J2215+5135 \protect\cite{Linares2018} and PSR J095-0607,
respectively. The red region amounts to the secondary component of GW190814.
The brown and the purple regions show two different reports of the pulsar
J0030+0451 \protect\cite{Miller2019,Riley2019}. The light blue band denotes
the remnant mass of GW190425.}
\label{massnjl9}
\end{figure}

\begin{figure}[h]
\center{\includegraphics[width=8.5cm] {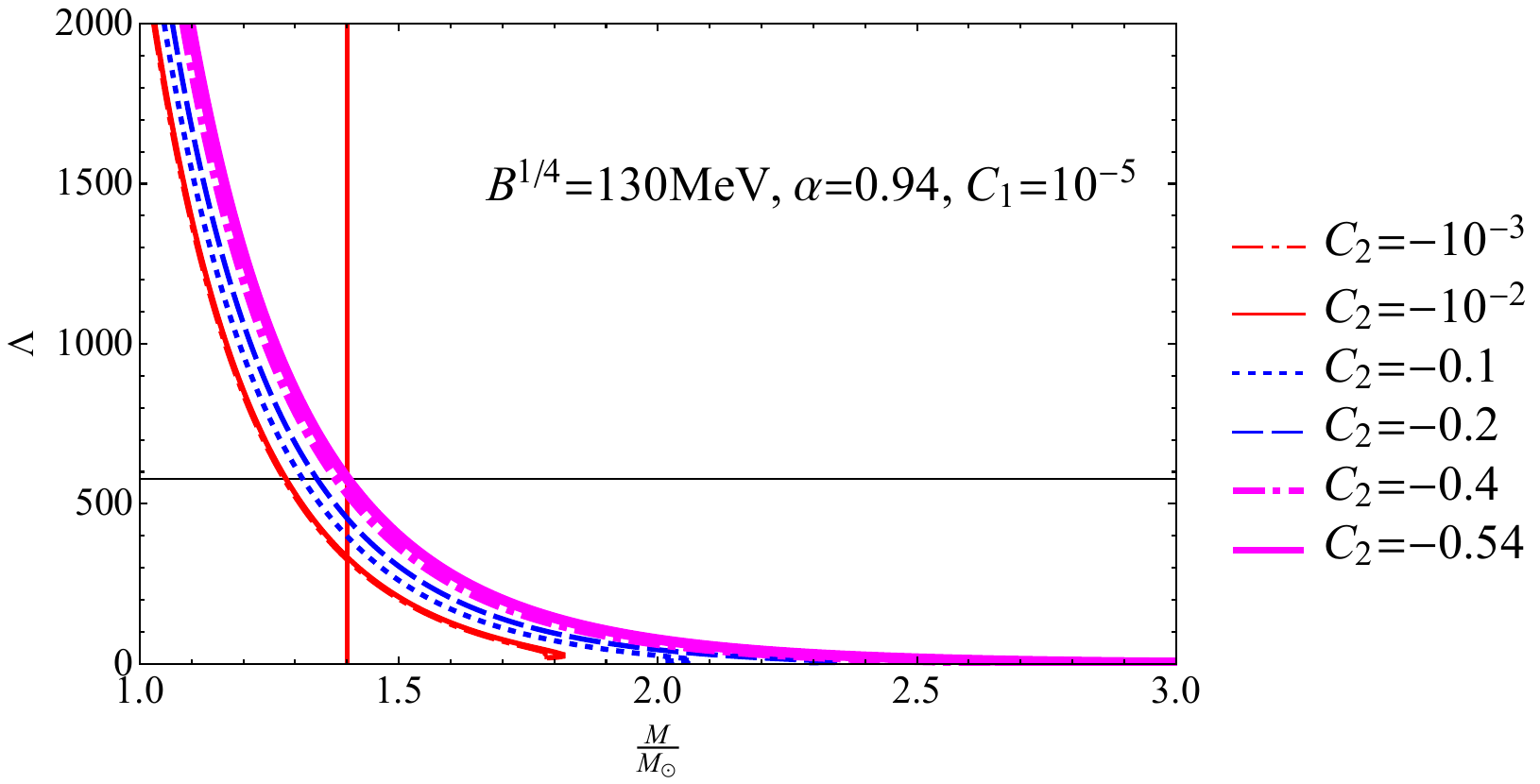}}
\caption{Dimensionless tidal deformability versus mass of the star for $%
B^{1/4}=130MeV$, $\protect\alpha =0.94$, $C_{1}=10^{-5}$, and different
values of $C_2$. The horizontal black line and the vertical red line
correspond to $\Lambda=580$ and $\frac{M}{{M}_{\odot }}=1.4$, respectively.}
\label{dataoftidal9}
\end{figure}

As the final investigation of SQS in dRGT-like massive gravity, we consider $%
B^{1/4}=130MeV$ and $\alpha=0.94$. Figures. \ref{massnjl9}, and \ref%
{dataoftidal9} show $M-R$ and $\Lambda-M$ diagrams for $C_{1}=10^{-5}$ and
different values of $C_2$, respectively. Figure \ref{dataoftidal9} shows
that for values $\arrowvert C_2\arrowvert \gtrsim 0.54$, the constraint $%
\Lambda_{1.4 {M}_{\odot }}\lesssim580$ is no longer satisfied. For this
reason, compared to the case $\alpha=0.8$, a smaller area of the $C_2$
parameter is accessible, which results from the sharpness of the EOS. The
structural properties of SQS for $B^{1/4}=130MeV$, $\alpha =0.94$, $%
C_{1}=10^{-5}$ and different values of $C_2$ are presented in Table. \ref%
{structural propertie9}.

Briefly, the maximum mass of SQSs with $B^{1/4}=130MeV$, and $\alpha=0.94$,
cannot be more than $3.45{M}_{\odot}$. However, the mentioned values of the
parameters of our system can satisfy the mass gap region (see Table. \ref%
{structural propertie9}, for more details). Besides, there are the same
behaviors of the compactness and the Schwarzschild radius when the value of $%
\lvert C_{2}\lvert$ increases. In other words, the strength of gravity
augments by increasing the value of $\lvert C_{2}\lvert$.
\begin{table*}[tbp]
\caption{Structural properties of SQS for $B^{1/4}=130MeV$, $\protect\alpha %
=0.94$, $C_{1}=10^{-5}$, and different values of $C_2$. The results falling
in mass gap region have been separated in a different column.}
\label{structural propertie9}\centering
\begin{tabular}{|c||c|c|c|c|c|c|c|c|}
\hline
\multicolumn{4}{c}{none mass gap} &  & \multicolumn{4}{c}{mass gap} \\ \hline
$C_2 $ & $-10^{-3}$ & $-10^{-2}$ & $-0.1$ & $-0.2$ & $-0.3$ & $-0.4$ & $%
-0.54 $ & $-0.55$ \\ \hline
$R(km)$ & $9.81$ & $9.85$ & $10.30$ & $10.77$ & $11.23$ & $11.63$ & $12.32$
& \hspace{0.1cm} $12.35$ \\ \hline
$M_{TOV}({M}_\odot)$ & $1.79 $ & $1.81$ & $2.06$ & $2.35$ & $2.65$ & $2.96$
& $3.42$ & \hspace{0.1cm} $3.45$ \\ \hline
$\Lambda_{1.4 {M}_{\odot }} $ & $326.43$ & $333.26$ & $398.93$ & $452.51$ & $%
499.87$ & $532.37$ & $578.95$ & \hspace{0.1cm} $580.64$ \\ \hline
$\Lambda_{M_{TOV}} $ & $32.25$ & $32.26$ & $15.20$ & $6.51$ & $2.92$ & $1.36$
& $0.35$ & \hspace{0.1cm} $0.32$ \\ \hline
$\sigma$ $(10^{-1})$ & $2.70$ & $2.72$ & $2.96$ & $3.23$ & $3.49$ & $3.77$ &
$4.11$ & \hspace{0.1cm} $4.13$ \\ \hline
$R_{Sch}(km) $ & $5.29 $ & $5.30$ & $5.54$ & $5.80$ & $6.03$ & $6.26$ & $%
6.57 $ & $6.59$ \\ \hline
\end{tabular}%
\end{table*}

As a final point, by considering the MNJL model with $B^{1/4}=130MeV$ and
different values of $\alpha$ ($0.5<\alpha<0.94$), we indicated that the mass
of SQSs could be in the mass gap region. So, it seems reasonable to consider
SQSs in massive gravity as candidates for the mass gap region.

\section{Conclusion}

We investigated the structural properties of quark star in massive gravity.
In this study, we considered three-flavor quark matter (up, down and strange
flavors) in MNJL model \cite{ChengMingLi2020} which is a combination of NJL
Lagrangian and its Fierz transformation by using weighting factors ($%
1-\alpha $) and $\alpha $. By imposing charge neutrality and chemical
equilibrium, we obtained the EOSs of SQM for three different values of $%
\alpha $, ($\alpha =0.5$, $\alpha =0.8$, and $\alpha =0.94$) and two
different values of Bag constant ($B^{1/4}=117MeV$ and $B^{1/4}=130MeV$). We
ensured that the EOSs satisfied causality and dynamical stability
conditions. We then presented a brief introduction on the modified TOV
equation in massive gravity. We explained the parameters appeared in TOV
equation in massive gravity, including $C_{1}$ and $C_{2}$. We also derived
the differential equation for the metric function $H(r)$ in massive gravity
to calculate the tidal deformability.

To assess the structural properties of SQS, we initially considered $%
B^{1/4}=117MeV$ in the EOS and then by setting a fixed value for $C_{1}$
(the mass-radius relation is independent of $C_{1}$ \cite{Behzad2017}) and
selecting different values of $C_{2}$, we calculated the structural
properties of SQS. The results analyzed to ensure compliance with the
constraint $\Lambda _{1.4{M}_{\odot }}\lesssim 580$. We observed that
increasing the $\alpha $ parameter led to an increase in the maximum
gravitational mass but also resulted the aforementioned constraint to be
satisfied in a smaller range of massive gravity parameters. Consequently,
the allowed mass range of the star decreased. For the case of $%
B^{1/4}=117MeV $, we obtained the maximum allowed $M_{TOV}$ approximately $%
2.13M_{\odot }$.

Subsequently, we altered, the value of $B$ and considered $B^{1/4}=130MeV$.
This change resulted in a softer EOS and a broader range of massive
parameters satisfying the astronomical constraints. We obtained masses that
fall within the mass gap region ($M\sim (2.5-5)M_{\odot }$) that
simultaneously satisfied all the observational constraints. Moreover, we
calculated Schwarzschild radius to show that compact objects studied in this
paper were not black holes.

In general relativity, soft EOSs yield small masses for quark stars, while
stiff EOSs fail to satisfy the the $\Lambda $ constraint. However, our study
showed that massive gravity allows for the existence of quark stars with
higher masses than those predicted by general relativity, while still
satisfying the constraints imposed by gravitational wave observations.

\textbf{Finally in Appendix B, by considering a constant value of $\alpha$ ($\alpha=0.8$), we increased the value of the bag constant to show that our results in massive gravity still cover the mass gap region.}
\begin{acknowledgements}
B. Eslam Panah thanks the University of Mazandaran. The
University of Mazandaran has supported the work of B. Eslam Panah by title
"Evolution of the masses of celestial compact objects in various gravities".
\end{acknowledgements}

\pagebreak

\appendix
\section{}

To calculate the tidal deformability, we need to obtain the metric function $%
H(r)$, which satisfies the following equation
\begin{eqnarray}
\left(-\frac{6e^{2\lambda}}{r^2}-2(\Phi ')^2+2\Phi
''+\frac{3}{r}\lambda '\right.\nonumber\\
\left.+\frac{7}{r}\Phi '- 2 \Phi'
\lambda '+ \frac{f}{r}(\Phi '+\lambda ')\right)H\nonumber\\
+ \left(\frac{2}{r}+\Phi '-\lambda'\right)\beta + \frac{d\beta}{dr} =0
\label{eq:h} ,
\end{eqnarray}
Here $\beta =dH/dr$, $f$ corresponds to $d\epsilon /dp$. Also, $\Phi^{\prime }$ satisfies the following equation
\begin{equation}
\Phi^{\prime }=-\frac{p^{\prime }}{(\epsilon +p)},  \label{A2}
\end{equation}%
where $p^{\prime }$ satisfies TOV equation in massive gravity.
\begin{equation}
p^{\prime }=\frac{(p+\epsilon )B_{2}}{2rB_{1}},  \label{A3}
\end{equation}%
\begin{equation}
M^{\prime }\left( r\right) =4\pi r^{2}\epsilon,  \label{A4}
\end{equation}%
where $B_{1}=$ $\frac{C_{1}r}{2}+C_{2}-1+\frac{2M\left( r\right) }{r}$, and $%
B_{2}=8\pi pr^{2}-\frac{C_{1}r}{2}+\frac{2M\left( r\right) }{r}$.

By deriving relations (\ref{A2}) and (\ref{A3}) with respect to $r$ we get
\begin{equation}
\Phi ^{\prime \prime }=\frac{\left( 1+f\right) {p^{\prime }}^{2}-p^{\prime
\prime }(\epsilon +p)}{(\epsilon +p)^{2}},  \label{A5}
\end{equation}%
and
\begin{align}
p^{\prime \prime }& =\frac{\left( 1+f\right) B_{2}p^{\prime }}{2rB_{1}}
\notag \\
& +\frac{\left( 4\pi r^{2}\left( rp^{\prime }+\epsilon \right) +8\pi r^{2}p-%
\frac{C_{1}r}{4}-\frac{M}{r}\right) (p+\epsilon )}{r^{2}B_{1}}  \notag \\
& -\frac{\left( C_{1}r+C_{2}-1+8\pi r^{2}\epsilon \right) (p+\epsilon )B_{2}%
}{2r^{2}B_{1}^{2}}.  \label{eq:P2}
\end{align}

In Eq. (\ref{eq:h}), $\lambda ^{\prime }$ satisfies the following equation
\begin{equation}
\lambda ^{\prime }=\frac{s^{\prime }}{2s},  \label{A7}
\end{equation}%
where
\begin{equation}
s=\frac{-1}{B_{1}},  \label{A8}
\end{equation}%
and
\begin{equation}
s^{\prime }=\frac{\frac{C_{1}}{2}+\frac{2M^{\prime }\left( r\right) }{r}-%
\frac{2M\left( r\right) }{r^{2}}}{B_{1}^{2}}.  \label{A9}
\end{equation}

Now, for the convenience of calculations, we write equation (\ref{eq:h}) as
below
\begin{equation}
\left( A_{1}+A_{2}+A_{3}\right) H+A_{4}\beta+\frac{d\beta}{dr}=0,
\label{eq:h2}
\end{equation}%
where $A_{1}$, $A_{2}$, $A_{3}$, and $A_{4}$ are defined as follows.
\begin{equation}
A_{1}\equiv -\frac{6e^{2\lambda }}{r^{2}},  \label{A11}
\end{equation}%
\begin{equation}
A_{2}\equiv -2\left( \Phi ^{\prime }\right) ^{2}+2\Phi ^{\prime \prime }+%
\frac{3\lambda ^{\prime }}{r}+\frac{7\Phi ^{\prime }}{r}-2\lambda ^{\prime
}\Phi ^{\prime },  \label{A12}
\end{equation}%
\begin{equation}
A_{3}\equiv \frac{(\Phi ^{\prime }+\lambda ^{\prime })f}{r},  \label{A13}
\end{equation}%
and
\begin{equation}
A_{4}\equiv \frac{2}{r}+\Phi ^{\prime }-\lambda ^{\prime }.  \label{A14}
\end{equation}

Now using Eqs. (\ref{A2}) to (\ref{A9}), we obtain the following terms for
coefficients $A_{1}$ to $A_{4}$.
\begin{equation}
A_{1}=\frac{6}{r^{2}B_{1}},  \label{A15}
\end{equation}%
\begin{align}
A_{2}& =\frac{B_{1}}{2r^{2}B_{1}^{2}}\left\{ \frac{B_{2}\left(
B_{2}f+B_{3}\right) }{B_{1}}-7B_{2}+3B_{3}\right.  \notag \\
& +\frac{1}{B_{4}}\left[ \frac{(f+1)B_{2}}{\left( B_{2}+\frac{C_{1}r}{2}%
\right) ^{-1}}+\frac{2B_{1}\left( B_{2}+\frac{C_{1}r}{2}\right) }{%
B_{4}\left( \frac{2M\left( r\right) }{r}+8\pi r^{2}\epsilon -2\right) ^{-1}}%
\right.  \notag \\
& +16\pi r^{2}\left[ \epsilon \left( \frac{C_{1}r}{4}+C_{2}+4\pi
pr^{2}-1\right) \right.  \notag \\
& \left. \left. \left. +\frac{3p}{\left( \frac{5C_{1}r}{12}+C_{2}+\frac{4\pi
pr^{2}}{3}-1\right) ^{-1}}+\frac{M\left( r\right) (7p+3\epsilon )}{r}\right] %
\right] \right\} ,  \label{A16}
\end{align}%
\begin{equation}
A_{3}=\frac{4\pi (p+\epsilon )f}{B_{1}},  \label{A17}
\end{equation}%
\begin{equation}
A_{4}=\frac{2\left( \frac{3C_{1}r}{4}+2\pi r^{2}(\epsilon -p)+C_{2}-1\right)
+\frac{2M\left( r\right) }{r}}{rB_{1}}.  \label{A18}
\end{equation}%
where $B_{3}=\frac{2M\left( r\right) }{r}-\frac{C_{1}r}{2}-8\pi \epsilon
r^{2}$, and $B_{4}=1-\frac{2M\left( r\right) }{r}$.

To check the correctness of the calculations, we consider the case $%
C_{1}=C_{2}=0$ and show that Eq. (\ref{eq:h2}) converts to the equation
obtained in Einstein's gravity
\begin{align}
& \frac{20H}{B_{4}}\left( \pi \left( \epsilon +\frac{9p+(\epsilon +p)f}{10}%
\right) -\frac{3}{10r^{2}}-\frac{\left( B_{2}+\frac{C_{1}r}{2}\right) ^{2}}{%
20rB_{4}}\right)  \notag \\
& -\frac{\left( \frac{2M\left( r\right) }{r}+4\pi r^{2}(\epsilon
-p)-2\right) H}{r\left( 1-\frac{2M\left( r\right) }{r}\right) }^{\prime
}+H^{\prime \prime }=0.  \label{A19}
\end{align}

By considering $C_{1}=C_{2}=0$, Eqs. (\ref{A15}) to (\ref{A17}) are
converted to
\begin{equation}
A_{1}=\frac{-6}{r^{2}B_{4}},  \label{A20}
\end{equation}%
\begin{equation}
A_{2}=\frac{4\pi B_{4}(9p+5\epsilon )-\left( B_{2}+\frac{C_{1}r}{2}\right)
^{2}}{B_{4}^{2}},  \label{A21}
\end{equation}%
\begin{equation}
A_{3}=\frac{4\pi (p+\epsilon )f}{B_{4}}.  \label{A22}
\end{equation}

It is clear that the sum of Eqs. (\ref{A20}) to (\ref{A22}) gives the
coefficient of $H$ in equation (\ref{A19}). Furthermore, by considering $%
C_{1}=C_{2}=0$, Eq. (\ref{A17}) becomes
\begin{equation}
A_{4}=\frac{2\left( 1+2\pi r^{2}(p-\epsilon )-\frac{M\left( r\right) }{r}%
\right) }{rB_{4}},  \label{A23}
\end{equation}%
which is equivalent to the coefficient of function $H^{\prime }$ in equation
(\ref{A19}).

\section{}
In the previous sections, we have considered two different values for the bag constant, namely $B^{1/4}=117MeV$ and $B^{1/4}=130MeV$ (similar to Ref. \cite{ChengMingLi2020}). The reason for choosing these two values of $B$ is to compare the structural properties of the SQS in massive gravity (which we have done in this paper) with the results obtained in Einstein gravity with the same values of $B$ (which was done in Ref. \cite{ChengMingLi2020}). One question that might arise  is the effect of increasing the bag constant on the structural properties of SQS. In this appendix, we increase  $B$ and check whether the results cover the mass gap region. The determination of the range of $B$ depends on various factors, such as the model used to obtain the EOS, the number of quark flavors, the mass of the quarks, etc. If we take into account the constraint of tidal deformability, determination of the range of $B$ also depends on the gravity used. This is because to obtain $\Lambda$, not only the EOS (in which $B$ appears) is effective, but also the gravity used is important. For example in Ref. \cite{Stergioulas}, by using MIT bag model and considering $m_s=0$ ($m_s$ is the mass of strange quark), $B$ is constrained in the range $58.9 \frac{MeV}{fm^3}<B<91.5 \frac{MeV}{fm^3}$. In \cite{Farhi1984} by assuming $m_s=150MeV$, the range of $B$ is $56-78  \frac{MeV}{fm^3}$. In \cite{Aziz2019} by studying 20 compact stars, $B$ is obtained in the range $41.58-319.31\frac{MeV}{fm^3}$. In Ref. \cite{Zhou2018}, using perturbative QCD and the $\Lambda$ constraint from GW170817, $B^{\frac{1}{4}}$ is constrained to a narrow range from 134.1$MeV$ to 141.4 $MeV$ ($42\frac{MeV}{fm^3}<B<52\frac{MeV}{fm^3}$). In \cite{Song1992} $B$ is in the range $100-200MeV^4$ ($13\frac{MeV}{fm^3}<B<208\frac{MeV}{fm^3}$). In this paper, we increase the value of $B$ in the MNJL model for a fixed value of $\alpha$ (e.g. $\alpha=0.8$) and investigate its effect on the structural properties of SQS in massive gravity. In the following, we choose two different values for $B$ including $B^{\frac{1}{4}}=140MeV$ ($B=50\frac{MeV}{fm^3}$) and $B^{\frac{1}{4}}=148MeV$ ($B=62\frac{MeV}{fm^3}$) and check whether the  results cover the mass gap region. We show that for these values of $B$, the results fall within the mass gap region. We do not try the values $\alpha=0.8$, and $B^{\frac{1}{4}}>148MeV$, because for these values, the results  do not satisfy the constraint $70\lesssim\Lambda_{1.4 {M}_{\odot }}\lesssim580$.
\subsection{The results for $\alpha=0.8$ and $B^{\frac{1}{4}}=140MeV$}
\begin{figure}[h]
	\center{\includegraphics[width=9cm] {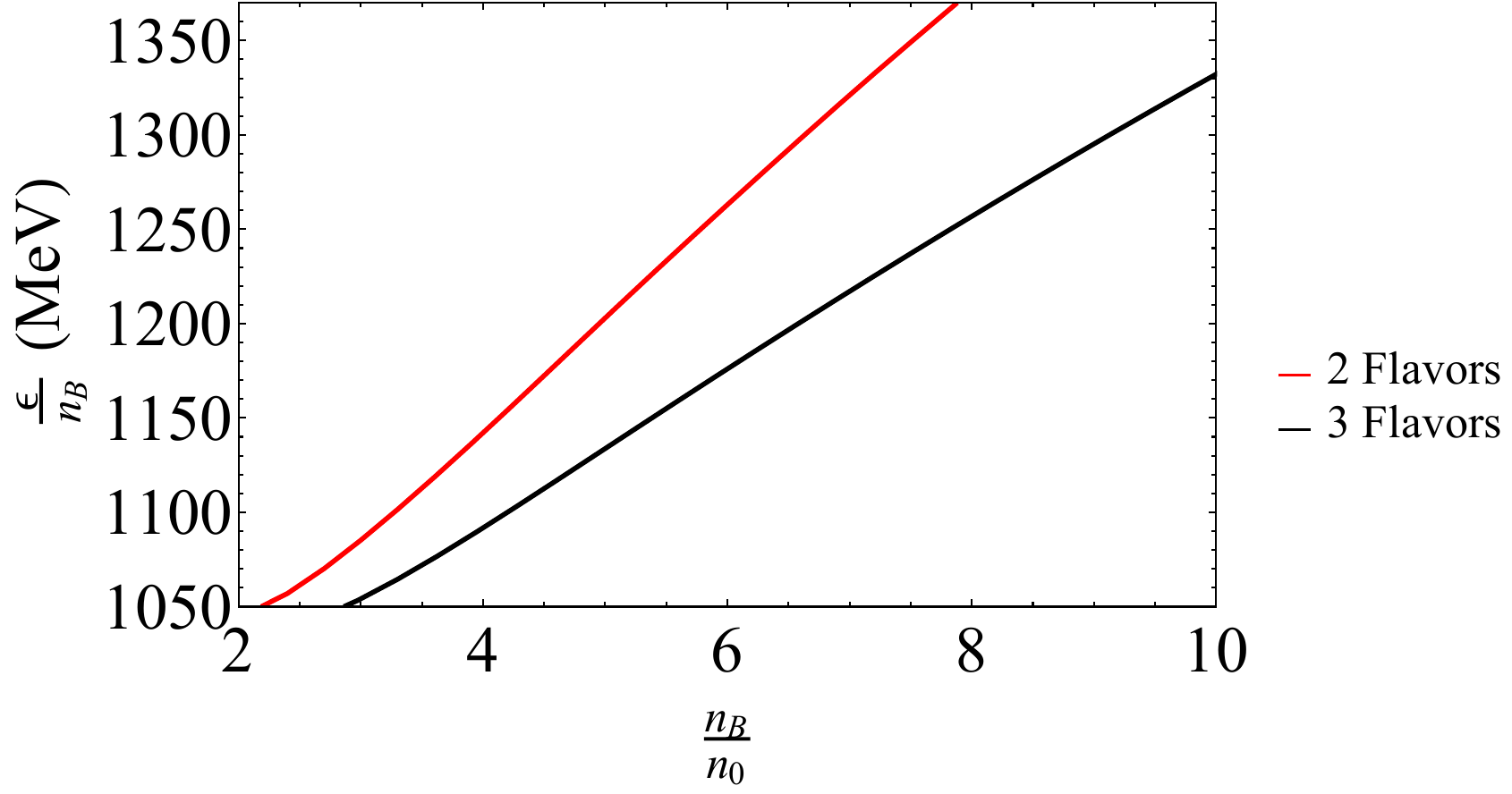}}
	\caption{Energy per baryon versus baryon number density for $B^{1/4}=140MeV$ and $\protect\alpha =0.8$. $n_0$ is the nuclear saturation density which is equal to $0.16 fm^{-3}$.}
	\label{stability condition}
\end{figure}

\begin{figure}[h]
	\center{\includegraphics[width=8.5cm] {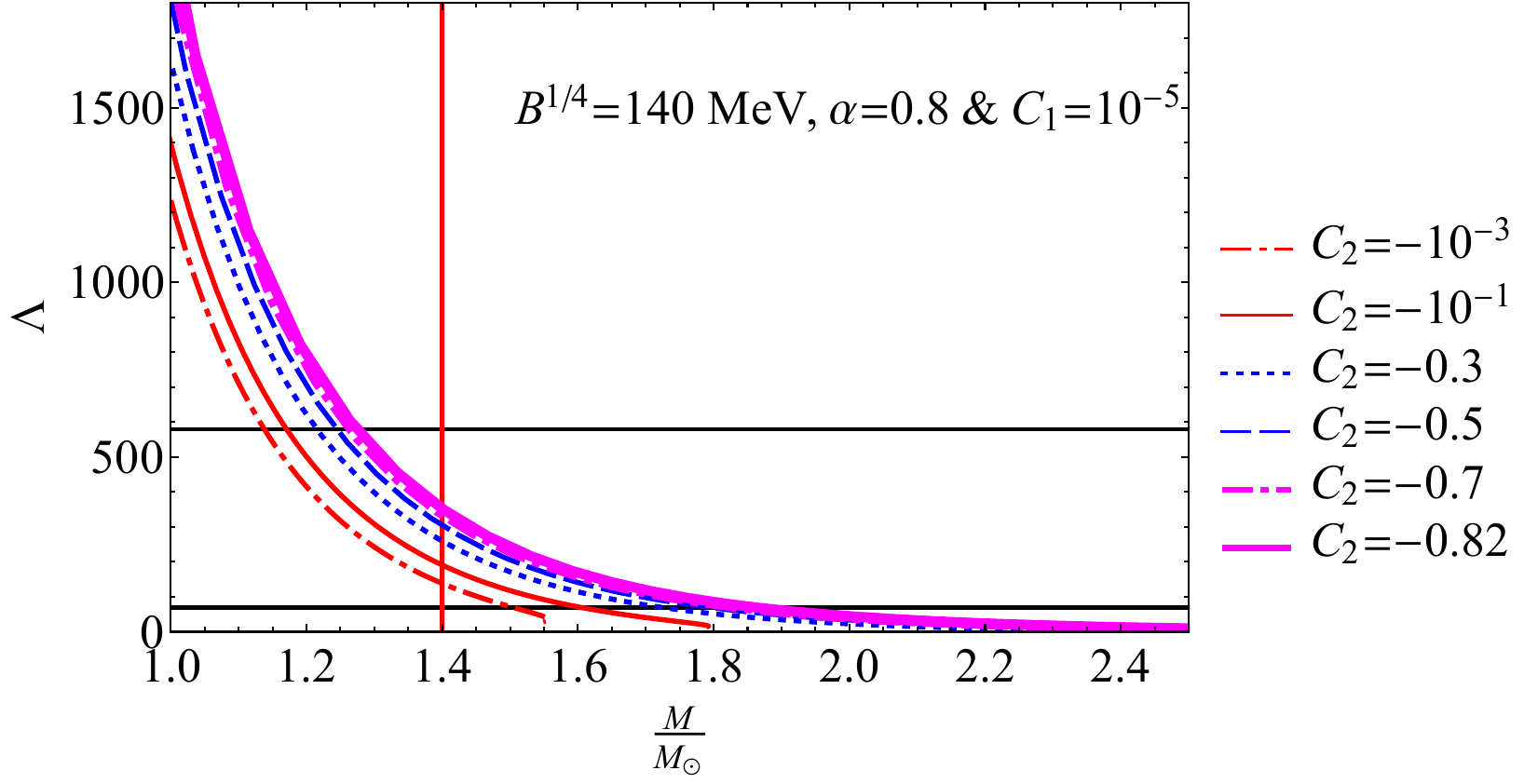}}
	\caption{Dimensionless tidal deformability ($\Lambda$) versus mass of the star for $B^{1/4}=140MeV$, $\protect\alpha =0.8$, $C_{1}=10^{-5}$, and different values of $C_2$.  The upper black line, the lower black line,  and the vertical red line correspond to $\Lambda=580$, $\Lambda=70$, and $\frac{M}{{M}_{\odot }}=1.4$, respectively.}
	\label{dataoftidal-140}
\end{figure}

Now we increase the value of $B^{\frac{1}{4}}$ to $140MeV$ and show that the results are still in the mass gap region. We first investigate the stability condition of SQM. Figure. \ref{stability condition} shows that the energy per baryon of three-flavor quark matter is less than that of two-flavor quark matter for $B^{1/4}=140MeV$ and $\protect\alpha =0.8$ in all baryon number densities. In the following, we investigate $\Lambda-M$ and $M-R$ diagrams for $B^{1/4}=140MeV$, $\protect\alpha =0.8$, $C_{1}=10^{-5}$, and different values of $C_2$ parameter. Figure. \ref{dataoftidal-140} shows that the condition $70\lesssim\Lambda_{1.4 {M}_{\odot }}\lesssim580$ is fulfilled for all values of $C_2$. Compared to the Fig. \ref{dataoftidal8} ($\alpha=0.8$ and $B^{1/4}=130MeV$), we can see that the values of $\Lambda$ decrease with the increase of $B$. This result is obvious, because as $B$ increases, the EOS becomes softer and consequently $\Lambda$ decreases.

\begin{figure}[h]
	\center{\includegraphics[width=9cm] {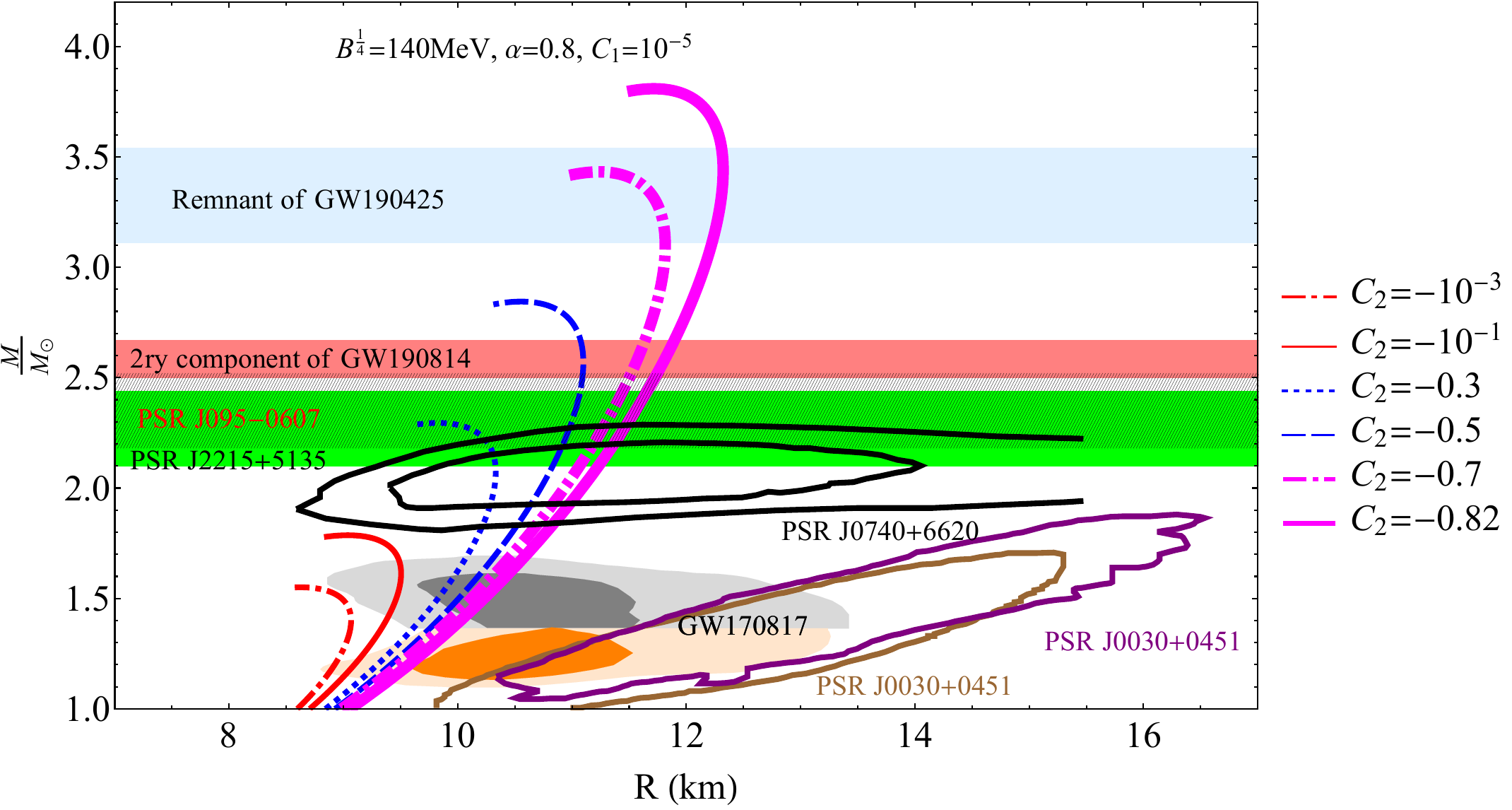}}
	\caption{Mass-Radius diagram for $B^{1/4}=140MeV$, $\alpha=0.8$, $C_{1}=10^{-5}$, and different values of $C_2$. The gray and orange regions are the mass - radius constraints
		from the GW170817 event. The black region shows pulsar J0740+6620, The green and black hatched regions represent pulsars J2215+5135 \cite{Linares2018} and PSR J095-0607, respectively. The red region amounts to the secondary component of GW190814. The brown and the purple regions show two different reports of the pulsar J0030+0451 \cite{Miller2019,Riley2019}. The light blue band denotes the remnant mass of GW190425.}
	\label{massnjl-140}
\end{figure}

Figure. \ref{massnjl-140} represents $M-R$ diagram for $B^{1/4}=140MeV$, $\protect\alpha =0.8$, $C_{1}=10^{-5}$, and different values of $C_2$. We can see that the results  still cover the mass gap region. However, compared to the Fig. \ref{massnjl8} ($\alpha=0.8$ and $B^{1/4}=130MeV$), the results are obtained in a narrower range of the mass gap. In fact, such a result can be expected by making the EOS softer.
\subsection{The results for $\alpha=0.8$ and $B^{\frac{1}{4}}=148MeV$}
\begin{figure}[h]
	\center{\includegraphics[width=9cm] {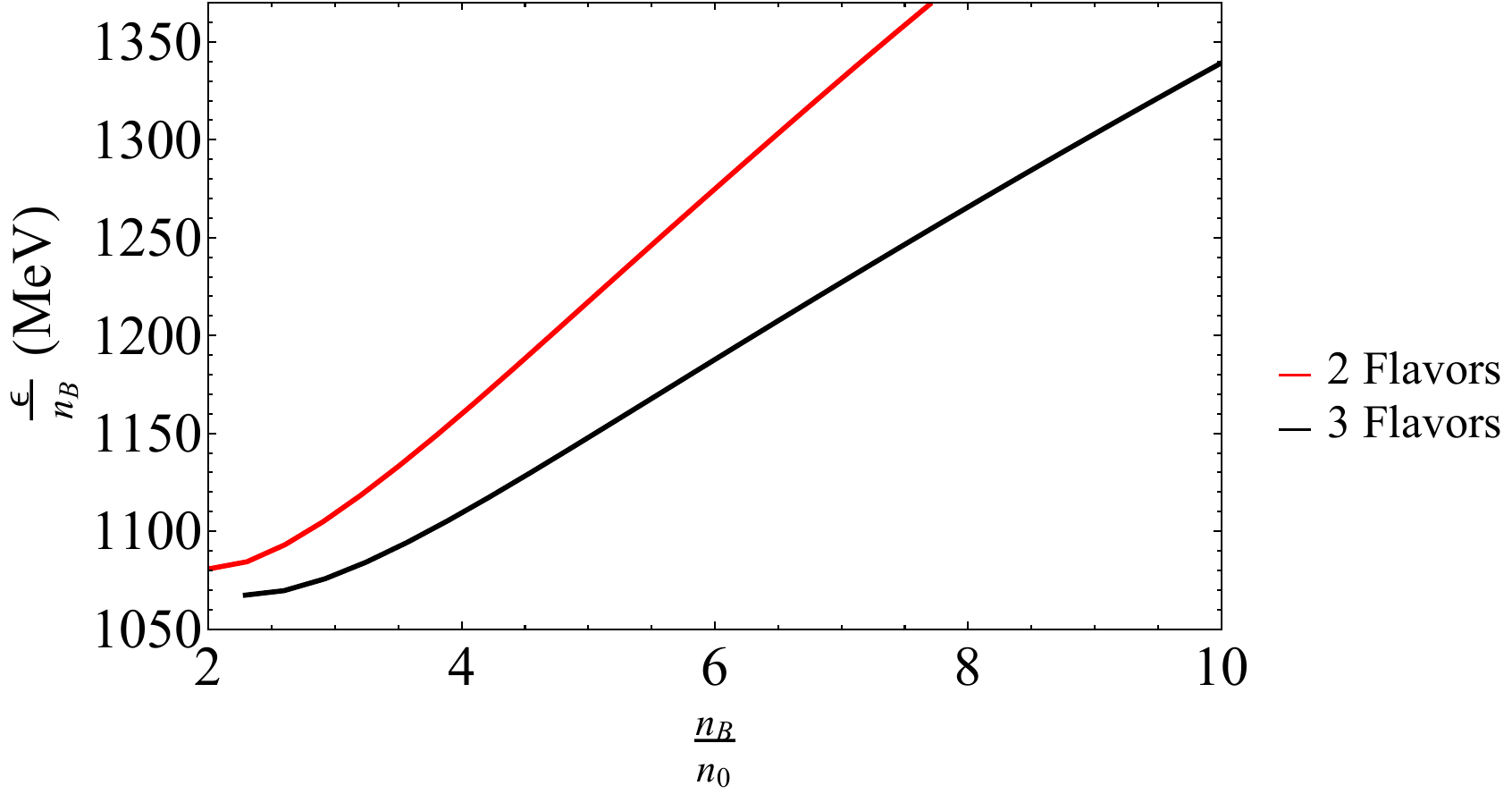}}
	\caption{Energy per baryon versus baryon number density for $B^{1/4}=148MeV$ and $\protect\alpha =0.8$. $n_0$ is the nuclear saturation density which is equal to $0.16 fm^{-3}$.}
	\label{stability condition2}
\end{figure}

\begin{figure}[h]
	\center{\includegraphics[width=8.5cm] {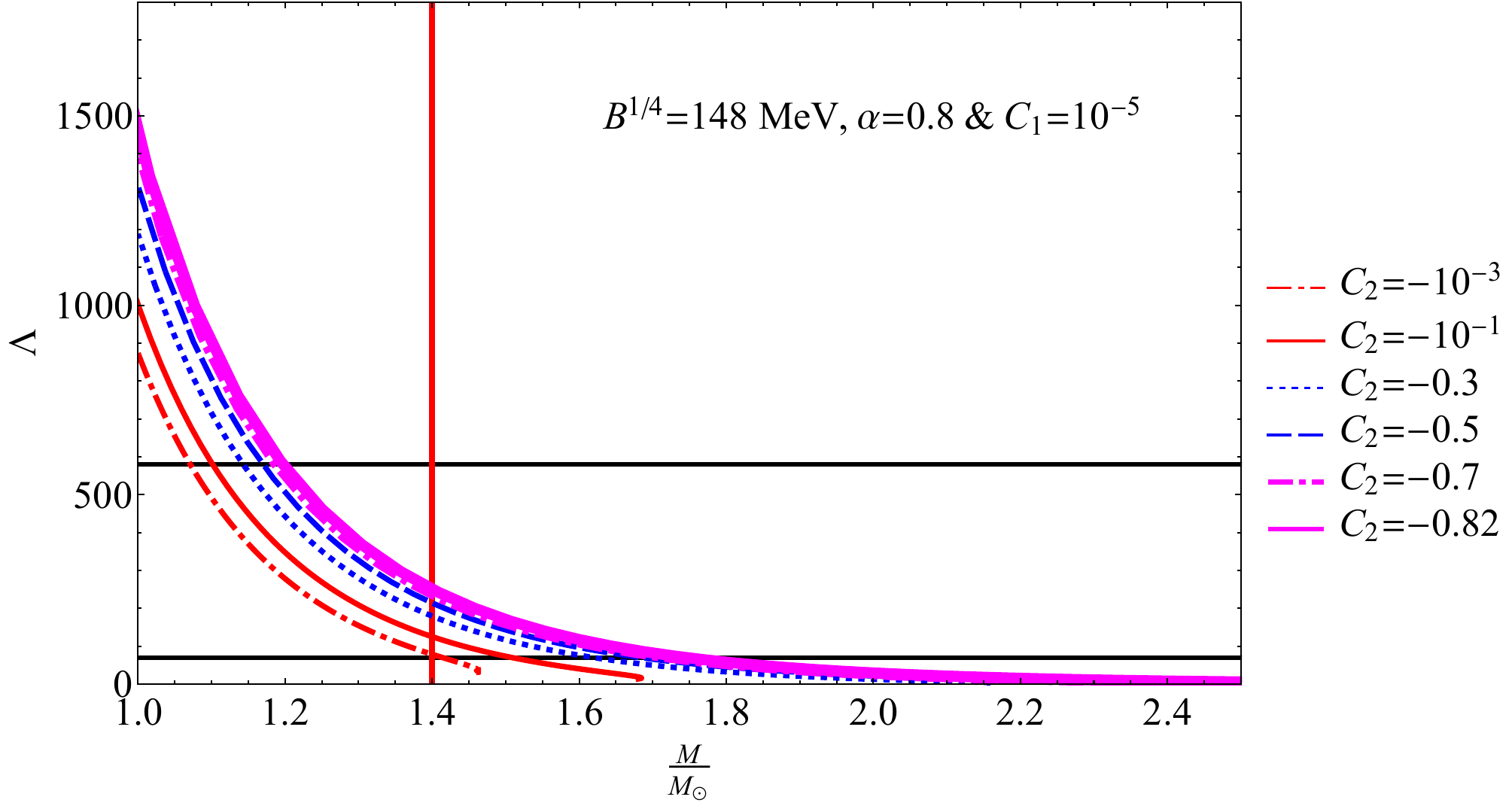}}
	\caption{Dimensionless tidal deformability ($\Lambda$) versus mass of the star for $B^{1/4}=148MeV$, $\protect\alpha =0.8$, $C_{1}=10^{-5}$, and different values of $C_2$.  The upper black line, the lower black line,  and the vertical red line correspond to $\Lambda=580$, $\Lambda=70$, and $\frac{M}{{M}_{\odot }}=1.4$, respectively.}
	\label{dataoftidal-148}
\end{figure}

\begin{figure}[h]
	\center{\includegraphics[width=9cm] {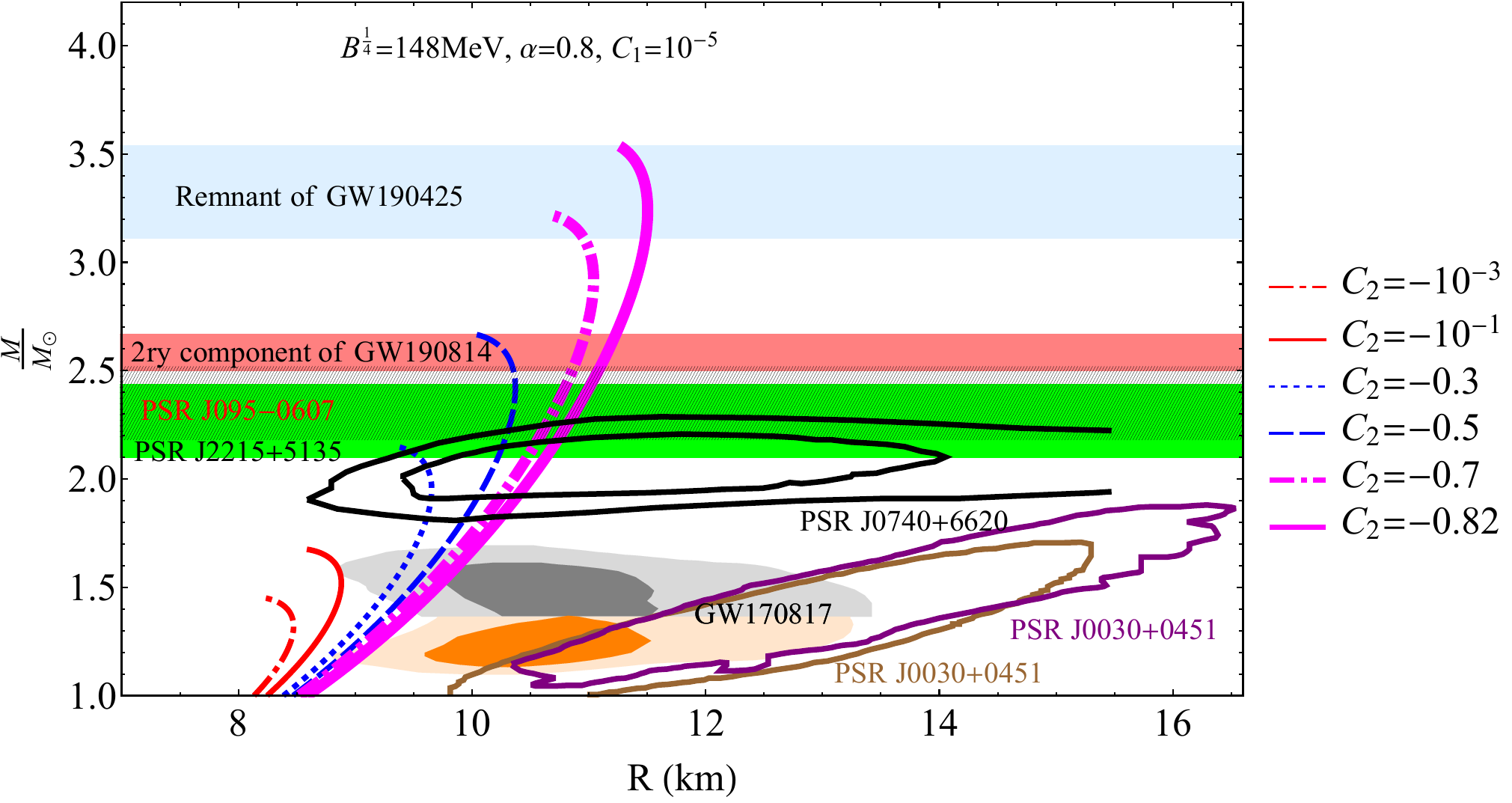}}
	\caption{Mass-Radius diagram for $B^{1/4}=148MeV$, $\alpha=0.8$, $C_{1}=10^{-5}$, and different values of $C_2$. The gray and orange regions are the mass - radius constraints
		from the GW170817 event. The black region shows pulsar J0740+6620, The green and black hatched regions represent pulsars J2215+5135 \cite{Linares2018} and PSR J095-0607, respectively. The red region amounts to the secondary component of GW190814. The brown and the purple regions show two different reports of the pulsar J0030+0451 \cite{Miller2019,Riley2019}. The light blue band denotes the remnant mass of GW190425.}
	\label{massnjl-148}
\end{figure}

For further investigation, we increase the value of $B^{1/4}$ to $148MeV$. First of all, we have to check the stability condition of the SQM. Figure. \ref{stability condition2} shows that this condition is well established  for all baryon number densities.
In the next step, we examine the $\Lambda$ behavior. Figure. \ref{dataoftidal-148} represents $\Lambda-M$ diagram for $B^{1/4}=148MeV$, $\protect\alpha =0.8$, $C_{1}=10^{-5}$, and different values of $C_2$. As we can see from this figure, the results respect to the constraint $70\lesssim\Lambda_{1.4 {M}_{\odot }}\lesssim580$. It is worth noting that for $\alpha=0.8$ and $B^{\frac{1}{4}}>148MeV$ the EOS becomes so soft that the $\Lambda_{1.4 {M}_{\odot }}$  is below $70$ for some values of $C_2$ and the constraint $70\lesssim\Lambda_{1.4 {M}_{\odot }}\lesssim580$ is lost. Therefore, we consider $B^{1/4}=148MeV (B=62\frac{MeV}{fm^3})$ as the maximum allowable value  of the bag constant in the  MNJL model for $\alpha=0.8$  to study the structural properties of SQM in massive gravity.
Figure. \ref{massnjl-148} shows $M-R$ diagram for $B^{1/4}=148MeV$, $\protect\alpha =0.8$, $C_{1}=10^{-5}$, and different values of $C_2$. As this figure shows, despite increasing $B$, the results still fall within the mass gap region. As mentioned above, this value of $B$ is the maximum possible value to meet $\Lambda_{1.4 {M}_{\odot }}$ constraint in MNJL for $\alpha=0.8$ in massive gravity.


\begin{thebibliography}{99}
\bibitem{Abbott2017} B.\thinspace P. Abbott, et al. (LIGO Scientific
Collaboration and Virgo Collaboration), Phys. Rev. Lett. \textbf{119},
161101 (2017).

\bibitem{Abbott2017b} M. Soares-Santos, et al., Astrophys. J. Lett. \textbf{%
\ 848}, L16 (2017).

\bibitem{Shibata2019} M. Shibata, E. Zhou, K. Kiuchi, and S. Fujibayashi,
Phys. Rev. D \textbf{100}, 023015 (2019).

\bibitem{Rezzolla2018} L. Rezzolla, E. R. Most, and L. R. Weih, Astrophys.
J. Lett. \textbf{852} L25 (2018).

\bibitem{J0952} R. W. Romani, et al., Astrophys. J. Lett. \textbf{934}, L17
(2022).

\bibitem{Abbott2020} R. Abbott, et al., Astrophys. J. Lett. \textbf{896},
L44 (2020).

\bibitem{ZMiao2021} Z. Miao, J. L. Jiang, A. Li, L.W. Chen, Astrophys. J.
Lett. \textbf{917}, L22 (2021).

\bibitem{Gao2020} H. Gao, et al., Front. Phys. \textbf{15}, 24603 (2020).

\bibitem{JSedaghat} J. Sedaghat, et al., Phys. Lett. B \textbf{833}, 137388
(2022).

\bibitem{ApJ499} C. D. Bailyn, et al., Astrophys. J. \textbf{499}, 367
(1998).

\bibitem{ApJ725} F. Özel, et al., Astrophys. J. \textbf{725}, 1918 (2010).

\bibitem{ApJ757} K. Belczynski, et al., Astrophys. J. \textbf{757}, 91
(2012).

\bibitem{ApJ941130} L. M. de Sá, et al., Astrophys. J. \textbf{941}, 130
(2022).

\bibitem{MNRAS5162022} A. Olejak, Ch. L. Fryer, K. Belczynski, and V.
Baibhav, Mon. Not. Roy. Astron. Soc. \textbf{516}, 2252 (2022).

\bibitem{Drozda2022} P. Drozda, et al., A\&A. \textbf{667}, 126 (2022).

\bibitem{IBombaci2021} I. Bombaci, et al., Phys. Rev. Lett. \textbf{126},
162702 (2021).

\bibitem{ITews2020} I. Tews, et al., Astrophys. J. Lett. \textbf{908}, L1
(2021).

\bibitem{ERMost2020} E. R. Most, L. J. Papenfort, L. R. Weih, and L.
Rezzolla, Mon. Not. Roy. Astron. Soc. Lett. \textbf{499}, L82 (2020).

\bibitem{NBZhang2020} N. -B. Zhang, and B. -A. Li, Astrophys. J. \textbf{902}%
, 38 (2020).

\bibitem{VDexheimer2020} V. Dexheimer, et al., Phys. Rev. C \textbf{103},
025808 (2021).

\bibitem{Blaschke2001} D. Blaschke, H. Grigorian, and D. Voskresensky,
Astron. Astrophys. \textbf{368}, 561 (2001).

\bibitem{Burgio2003} G. Burgio, H. J. Schulze, and F. Weber, Astron.
Astrophys. \textbf{408}, 675 (2003).

\bibitem{Alford2005} M. Alford, M. Braby, M. Paris, and S. Reddy, Astrophys.
J. \textbf{629}, 969 (2005).

\bibitem{Pal2023} S. Pal, S. Podder, D. Sen, and G. Chaudhuri, Phys. Rev. D
\textbf{107}, 063019 (2023).

\bibitem{Rather2023} I. A. Rather, et al., Astrophys. J. \textbf{943} 52
(2023).

\bibitem{Li2023} J. J. Li, A. Sedrakian, and M. Alford, Phys. Rev. D \textbf{%
107}, 023018 (2023).

\bibitem{Michel1988} F. C. Michel, Phys. Rev. Lett. \textbf{60}, 677 (1988).

\bibitem{Drago2001} A. Drago, and A. Lavagno, Phys. Lett. B \textbf{511},
229 (2001).

\bibitem{Kurkela2010} A. Kurkela, P. Romatschke, and A. Vuorinen, Phys. Rev.
D \textbf{81}, 105021 (2010).

\bibitem{Wang2019} Q. Wang, C. Shi, and H. -S. Zong, Phys. Rev. D \textbf{100%
}, 123003 (2019).

\bibitem{Deb2021} D. Deb, B. Mukhopadhyay, and F. Weber, Astrophys. J.
\textbf{922}, 149 (2021).

\bibitem{Terazawa1989} H. Terazawa, and J. Phys. Soc. Jpn. \textbf{58}, 3555
(1989).

\bibitem{Witten1984} E. Witten, Phys. Rev. D \textbf{30}, 272 (1984).

\bibitem{ABodmer} A. Bodmer, Phys. Rev. D \textbf{4}, 1601 (1971).

\bibitem{Daniel2021} D. A. Godzieba, et al., Astrophys. J. \textbf{908}, 122
(2021).

\bibitem{AngLi2021} A. Li, et al., Astrophys. J. \textbf{913}, 27 (2021).

\bibitem{Ferreira2021} M. Ferreira, R. C. Pereira, and C. Providência, Phys.
Rev. D \textbf{103}, 123020 (2021).

\bibitem{Miller2019} M. C. Miller, et al., Astrophys. J. Lett. \textbf{887},
L24 (2019).

\bibitem{Riley2019} T. E. Riley, et al., Astrophys. J. Lett. \textbf{887},
L21 (2019).

\bibitem{ShuHua2020} S. -H. Yang, et al., Astrophys. J. \textbf{902}, 32
(2020).

\bibitem{ALi2021} A. Li, et al., Mon. Not. Roy. Astron. Soc. \textbf{506},
5916 (2021).

\bibitem{TERiley2019} T. E. Riley, et al., Astrophys. J. Lett. \textbf{887},
L21 (2019).

\bibitem{Roupas2021} Z. Roupas, G. Panotopoulos, and I. Lopes, Phys. Rev. D
\textbf{103}, 083015 (2021).

\bibitem{Lopes2021} L. L. Lopes, and D. P. Menezes, Astrophys. J. \textbf{936%
}, 41 (2022).

\bibitem{CZhang2021} C. Zhang, Phys. Rev. D \textbf{104}, 083032 (2021).

\bibitem{MassI} K. Koyama, G. Niz, and G. Tasinato, Phys. Rev. Lett. \textbf{%
107}, 131101 (2011).

\bibitem{MassII} A. H. Chamseddine, and M. S. Volkov, Phys. Lett. B \textbf{%
704}, 652 (2011).

\bibitem{MassIII} G. D'Amico, C. de Rham, S. Dubovsky, G. Gabadadze, D.
Pirtskhalava, and A. J. Tolley, Phys. Rev. D \textbf{84}, 124046 (2011).

\bibitem{MassIV} K. Hinterbichler, Rev. Mod. Phys. \textbf{84}, 671 (2012).

\bibitem{MassV} Y. Akrami, T. S. Koivisto, and M. Sandstad, JHEP. \textbf{03}%
, 099 (2013).

\bibitem{Zhang2019} J. Zhang, and S.-Y. Zhou, Phys. Rev. D \textbf{97},
081501(R) (2018).

\bibitem{LIGOI} B. P. Abbott, et al. (LIGO Scientific and Virgo
Collaborations), Phys. Rev. Lett. \textbf{116}, 061102 (2016).

\bibitem{LIGOII} B. P. Abbott, et al. (LIGO Scientific and Virgo
Collaborations), Phys. Rev. Lett. \textbf{116}, 221101 (2016).

\bibitem{massgraviton1} A. S. Goldhaber, and M. M. Nieto, Rev. Mod. Phys.
\textbf{82}, 939 (2010).

\bibitem{massgraviton2} J. B. Jimenez, F. Piazza, and H. Velten, Phys. Rev.
Lett. \textbf{116}, 061101 (2016).

\bibitem{massgraviton3} A.F. Ali, and S. Das, Int. J. Mod. Phys. D \textbf{25%
}, 1644001 (2016).

\bibitem{massgraviton4} C. de Rham, J. T. Deskins, A. J. Tolley, and S.-Y.
Zhou, Rev. Mod. Phys. \textbf{89}, 025004 (2017).

\bibitem{Kareeso2018} P. Kareeso, P. Burikham, and T. Harko, Eur. Phys. J. C
\textbf{78}, 941 (2018).

\bibitem{Yamazaki2019} M. Yamazaki, T. Katsuragawa, S. D. Odintsov, and S.
Nojiri, Phys. Rev. D \textbf{100}, 084060 (2019).

\bibitem{Katsuragawa2016} T. Katsuragawa, S. Nojiri, S. D. Odintsov, and M.
Yamazaki, Phys. Rev. D \textbf{93}, 124013 (2016).

\bibitem{Behzad2017} S. H. Hendi, G. H. Bordbar, B. Eslam Panah, S.
Panahiyan, JCAP \textbf{07}, 004 (2017).

\bibitem{Behzad2019} B. Eslam Panah, and H. L. Liu, Phys. Rev. D \textbf{99}%
, 104074 (2019).

\bibitem{deRham} C. de Rham, G. Gabadadze, and A. J. Tolley, Phys. Rev.
Lett. \textbf{106}, 231101 (2011).

\bibitem{Hinterbichler2012} K. Hinterbichler, Rev. Mod. Phys. \textbf{84},
671 (2012).

\bibitem{Hassan2012} S. F. Hassan, R. A. Rosen, and A. Schmidt-May, J. High
Energy Phys. \textbf{1202}, 026 (2012)

\bibitem{Vegh} D. Vegh, [arXiv:1301.0537].

\bibitem{Abbott2018} B.\thinspace P. Abbott et al. (LIGO Scientific
Collaboration and Virgo Collaboration), Phys. Rev. Lett. \textbf{121},
161101 (2018).

\bibitem{ChengMingLi2020} C. -M. Li, et al., Phys. Rev. D \textbf{101},
063023 (2020).

\bibitem{Hatsuda1985} T. Hatsuda, and T. Kunihiro, Prog. Theor. Phys.
\textbf{74}, 765 (1985).

\bibitem{weber 2005} F. Weber, Prog. Part. Nucl. Phys. \textbf{54}, 193
(2005).

\bibitem{Amico} G. D'Amico, C. de Rham, S. Dubovsky, G. Gabadadze, D.
Pirtskhalava, and A. J. Tolley, Phys. Rev. D \textbf{84}, 124046 (2011).

\bibitem{Akrami} Y. Akrami, T. S. Koivisto, and M. Sandstad, JHEP \textbf{03}%
, 099 (2013).

\bibitem{Berezhiani} L. Berezhiani, G. Chkareuli, C. de Rham, G. Gabadadze,
and A. J. Tolley. Phys. Rev. D \textbf{85}, 044024 (2011).

\bibitem{Knutsen1988} H. Knutsen, Mon. Not. Roy. Astron. Soc. \textbf{232},
163 (1988).

\bibitem{Mak2013} M. K. Mak, and T. Harko, Eur. Phys. J. C \textbf{73}, 2585
(2013).

\bibitem{Hendi2016} S. H. Hendi, G. H. Bordbar, B. Eslam Panah, and S.
Panahiyan, JCAP \textbf{09}, 013 (2016).

\bibitem{Linares2018} M. Linares, T. Shahbaz, and J. Casares, Astrophys. J.
\textbf{859}, 54 (2018).

\bibitem{Sedaghat20221} J. Sedaghat, S. M. Zebarjad, G. H. Bordbar, and B.
Eslam Panah, Phys. Lett. B \textbf{829}, 137032 (2022).

\bibitem{Cromartie2019} H. T. Cromartie, et al., Nat. Astron. \textbf{4}, 72
(2019).

\bibitem{Chia2021} H. Sh. Chia, Phys. Rev. D \textbf{104}, 024013 (2021).

\bibitem{Stergioulas} N. Stergioulas, Living Rev. Rel \textbf{6}, 3 (2013).

\bibitem{Farhi1984} E. Farhi, and R. L. Jaffe, Phys. Rev. D \textbf{30}, 2379 (1984).

\bibitem{Aziz2019} A. Aziz et al., Int. J. Mod. Phys. D \textbf{28}, 13 (2019).

\bibitem{Zhou2018} E. Zhou, X. Zhou, and  A. Li, Phys. Rev. D \textbf{97}, 083015 (2018).

\bibitem{Song1992} G. Song, W. Enke, and L. Jiarong,, Phys. Rev. D \textbf{46}, 3211 (1992).
\end{thebibliography}
\end{document}